\documentclass[aps,prd,preprint,groupedaddress]{revtex4}

\usepackage{graphicx}
\usepackage{bm}

\begin{document}



\title{Baryon Asymmetry of the Universe without Boltzmann or
Kadanoff-Baym}


\author{Jean-S\'{e}bastien Gagnon}
\email{jean-sebastien.gagnon@physik.tu-darmstadt.de}
\affiliation{\'{E}cole Polytechnique F\'{e}d\'{e}rale de Lausanne, Lausanne, Switzerland}
\affiliation{Technische Universit\"{a}t Darmstadt, Darmstadt, Germany}
\author{Mikhail Shaposhnikov}
\email{mikhail.shaposhnikov@epfl.ch}
\affiliation{\'{E}cole Polytechnique F\'{e}d\'{e}rale de Lausanne, Lausanne, Switzerland }


\date{\today}

\begin{abstract}

We present a formalism that allows the computation of the baryon
asymmetry of the universe from first principles of statistical physics
and quantum field theory that is applicable to certain types of beyond
the Standard Model physics (such as the neutrino Minimal Standard
Model -- $\nu$MSM) and does not require the solution of Boltzmann or
Kadanoff-Baym equations. The formalism works if a thermal bath of
Standard Model particles is very weakly coupled to a new sector
(sterile neutrinos in the $\nu$MSM case) that is out-of-equilibrium. 
The key point that allows a computation without kinetic equations is
that the number of sterile neutrinos produced during the relevant
cosmological period remains small.  In such a case, it is possible to
expand the formal solution of the von Neumann equation perturbatively
and obtain a master formula for the lepton asymmetry expressed in
terms of non-equilibrium Wightman functions.  The master formula
neatly separates $CP$-violating contributions from finite temperature
correlation functions and satisfies all three Sakharov conditions. 
These correlation functions can then be evaluated perturbatively; the
validity of the perturbative expansion depends on the parameters of
the model considered.  Here we choose a toy model (containing only two
active and two sterile neutrinos) to illustrate the use of the
formalism, but it could be applied to other models.

\end{abstract}


\maketitle


\section{Introduction}

The baryon asymmetry of the universe (BAU) is a very important
quantity in cosmology.  It is quantified by the (dimensionless)
baryon-to-photon ratio $\eta$.  This ratio has been measured very
precisely by the WMAP collaboration~\cite{WMAP_2008} and is given by
$\eta = (n_{b}-n_{\bar{b}})/n_{\gamma} = (6.1 \pm 0.2)\times
10^{-10}$.  Here $n_{b}$ ($n_{\bar{b}}$) is the number density of
baryons (anti-baryons) and $n_{\gamma}$ is the number density of
photons.  If this excess of baryons over anti-baryons is not simply an
initial condition of our universe, then it is the goal of any particle
physics models to explain this asymmetry.

In principle, the Standard Model possesses all the necessary
ingredients to produce an asymmetry (i.e. it fulfills the three
Sakharov conditions~\cite{Sakharov_1967}): baryon number violating
processes are mediated by sphalerons, $CP$-violation is hidden in the
Cabibbo Kobayashi Maskawa (CKM) matrix and a first order phase
transition would provide the necessary out-of-equilibrium condition. 
On the other hand, it has been shown that in the Standard Model there
is no first order electroweak phase transition for Higgs masses above
80 GeV (it is a smooth crossover)~\cite{Kajantie_etal_1996}.  Since
the lower bound on the Higgs mass from LEP is 114 GeV, the
out-of-equilibrium condition necessary for baryogenesis is not
satisfied and no asymmetry is produced.  In extensions of the Standard
Model, this conclusion may change; thus new physics beyond the
Standard Model is necessary to explain the BAU.

There exists many mechanisms/models that could explain the BAU.  The
most well-known are Grand Unified Theories (GUT) baryogenesis,
electroweak baryogenesis in extensions of the Standard Model,
leptogenesis and the Affleck-Dine mechanism.  See
Refs.~\cite{Trodden_1998,Riotto_Trodden_1999,Dine_Kusenko_2004,Buchmuller_2007,Davidson_etal_2008,Pilaftsis_2009,Shaposhnikov_2009}
for reviews.  A common feature of these models is that they require
heavy degrees of freedom that are hard to detect with present
accelerators (the LHC may discover some of these particles in the near
future).

An alternative model is the neutrino Minimal Standard Model
($\nu$MSM)~\cite{Asaka_etal_2005,Asaka_Shaposhnikov_2005,
Asaka_etal_2006,Asaka_etal_2006a,Bezrukov_Shaposhnikov_2007,
Laine_Shaposhnikov_2008,Shaposhnikov_2008,Roy_Shaposhnikov_2010,
Canetti_Shaposhnikov_2010};
it is a minimal extension of the Standard Model with three sterile
right-handed neutrinos with masses below the electroweak scale.  This
model has interesting features.  For instance, it could explain
simultaneously three shortcomings of the Standard Model (namely
neutrino oscillations, dark matter and BAU).  Also, since it contains
no very heavy degree of freedom, it could in principle be tested
experimentally with present facilities \cite{Gorbunov_Shaposhnikov_2007,Shaposhnikov_2008a}.

For baryogenesis to occur some degrees of freedom must be out of
thermal equilibrium in order to fulfill the third Sakharov condition. 
The usual approach to baryon excess computations is to use Boltzmann
equations.  Various assumptions are used in the derivation of
Boltzmann equations, one of them being that the coherence length of
the processes involved must be much smaller than the mean free path of
the particles.  As long as these assumptions are satisfied, Boltzmann
equations can describe systems that are arbitrarily
out-of-equilibrium.  But it is shown in
Refs.~\cite{Buchmuller_Fredenhagen_2000,Lindner_Muller_2006} that
coherence effects are important in baryogenesis, and thus a more
refined quantum mechanical treatment is needed.

Out-of-equilibrium quantum field theory is a notoriously difficult
subject.  Significant progress has been made recently in formal
aspects (e.g.~\cite{Berges_2001,Aarts_etal_2002,Berges_2004}) but
applications to realistic baryogenesis computations are still lacking
(although see the recent progress in
Refs.~\cite{DeSimone_Riotto_2007,Anisimov_etal_2008,
Cirigliano_etal_2009,Garny_etal_2009a,Garny_etal_2009b,Beneke_etal_2010}).  

Our utlimate goal is to compute the BAU in the phenomenologically
interesting $\nu$MSM.  Previous calculations of the BAU in the
$\nu$MSM \cite{Asaka_Shaposhnikov_2005} show that the produced
asymmetry is in the right range.  The production of a baryon asymmetry
in the $\nu$MSM happens via coherent active-sterile neutrino
oscillations and requires appropriate kinetic equations for its
treatment.  These calculations are performed using kinetic equations
of the form
\cite{Sigl_Raffelt_1993,Akhmedov_etal_1998,Asaka_Shaposhnikov_2005,
Shaposhnikov_2008}:
\begin{eqnarray}
\label{eq:Kinetic_equation}
i\frac{d\rho}{dt} & = & \left[H,\rho\right]  -\frac{i}{2}\{\Gamma,\rho\} 
+ \frac{i}{2}\{\Gamma_{p},1-\rho\},
\end{eqnarray}
where $\rho$ is the complete neutrino density matrix, $H$ is the
Hamiltonian and $\Gamma_{p}$ and $\Gamma$ are production and
destruction rates, respectively.  The approach of
Ref.~\cite{Asaka_Shaposhnikov_2005} based on
Eq.~(\ref{eq:Kinetic_equation}) has several weak points.  First,
Eq.~(\ref{eq:Kinetic_equation}) relies on the usual assumptions of
kinetic theory (with the additional assumption that the duration of a
collision is small compared to the various oscillation times of
$\rho$).  Second, the calculations in
Ref.~\cite{Asaka_Shaposhnikov_2005} are done in the relaxation time
approximation and it is assumed that the integrals in the collision
term are dominated by $O(T)$ momenta.  This assumption might not be
warranted; since this is basically an oscillation problem with many
particles, there are many timescales involved and all components might
not relax in the same way.  Another weak point of
Eq.~(\ref{eq:Kinetic_equation}) is that it is not systematically
derived from first principles and thus there is no real control over
the error.  This has important consequences for phenomenology since
``factors of a few'' in the determination of the allowed parameter
range of the $\nu$MSM are crucial for the experimental searches of its
particles.  Thus in order to better constrain the model and study its
phenomenological implications, we need a first principles calculation
based on quantum field theory.  The first principles formalism
presented here and the treatment based on
Eq.~(\ref{eq:Kinetic_equation}) could also be compared in a region of
parameter space where the two approaches should match in order to
estimate the accuracy of kinetic theory approaches.

In this paper, we present the first steps toward a computation of the
baryon asymmetry from first principles of statistical physics and
quantum field theory.  We focus on a particular class of models
similar to the $\nu$MSM, namely a minimal extension of the Standard
Model with an arbitrary number of sterile neutrinos.  Our approach is
sufficiently general that it could be applied to other models.  The
baryogenesis scenario studied here is similar in spirit to
leptogenesis (with crucial differences).  The key idea is that in some
region of parameter space where the Yukawa couplings of the sterile
neutrinos are small, it is possible to use conventional perturbation
theory without having recourse to more sophisticated non-equilibrium
tools (such as Kadanoff-Baym equations). For a similar treatment used
in the context of sterile neutrino production, see Ref.~\cite{Asaka_etal_2006}.

The rest of the paper is organized as follows.  In
Sect.~\ref{Sec:Theoretical_background}, we present the model
Lagrangian used in this study and outline our baryogenesis scenario. 
The core of the paper is Sect.~\ref{Sec:Perturbative_formula}, where
we derive a perturbative formula that expresses the (lepton) asymmetry
in terms of Wightman functions and discuss its range of validity.  In
Sect.~\ref{Sec:Application}, we test the formula derived in
Sect.~\ref{Sec:Perturbative_formula} by applying it to a toy model
that can be solved both perturbatively and exactly.  This computation
also serves as an illustration of the inner workings of the
perturbative formula.  Some elements of this paper have already been
published (in a different form) in the proceedings of Strong and
Electroweak Matter 2008~\cite{Gagnon_2009}.

\section{Theoretical Background}
\label{Sec:Theoretical_background}

\subsection{Model Lagrangian}
\label{Sec:Model_Lagrangian}
A generic Lagrangian containing A active neutrinos and B sterile
(right handed) neutrinos can be written in the chiral basis as
($\alpha$ runs from $1$ to $A$ and $I$ from $1$ to $B$):
\begin{eqnarray}
\label{eq:Lagrangian}
{\cal L}_{AB} & = & \bar{L}_{\alpha}iD\!\!\!/ L_{\alpha} + 
\bar{N}_{I}i\partial\!\!\!/ N_{I} - \frac{M_{IJ}}{2}\bar{N}_{I}^{c}N_{J} 
- F_{\alpha I}\bar{L}_{\alpha}N_{I}\tilde{\Phi} + \mbox{h.c.},
\end{eqnarray}
where $L$ is the active lepton doublet, $N$ the sterile neutrino
singlet, $N^{c}$ the charge conjugated $N$, $\Phi$ the Higgs doublet,
$\rho$ the Higgs field, $M_{IJ}$ are Majorana masses and $F_{\alpha
I}$ are Yukawa couplings.  Here sterile means singlet under the
Standard Model gauge group.  Note that for the purpose of deriving a
master formula for the asymmetry, only active-sterile transitions are
necessary and other Standard Model particles are left out here.  The
case $A = 3$ and $B = 3$ (and including Standard Model particles)
corresponds to the $\nu$MSM.  Note that in the phenomenologically
relevant region of the $\nu$MSM parameter space where the right
abundance of dark matter is obtained, the ``dark matter sterile
neutrino'' has a tiny Yukawa coupling and essentially decouples
\cite{Asaka_etal_2005,Asaka_Shaposhnikov_2005}.  Thus the $\nu$MSM has effectively
three active and two sterile neutrinos.

We can check that the Lagrangian~(\ref{eq:Lagrangian}) satisfies all
three Sakarov's conditions for baryogenesis.  First, the presence of
the Majorana mass term breaks lepton number conservation (and thus
baryon number conservation via sphaleron processes).  Second, if
$F_{\alpha I}$ is complex, then there can be $CP$ violation in the
system (similar to the case of the CKM matrix). 
Table~(\ref{Table:Parameters}) shows the counting of physical
parameters (masses, Yukawa couplings, mixing angles and comlex phase)
in~(\ref{eq:Lagrangian}) for various numbers of active and sterile
neutrinos.  Note that complex phases can only be present for $B \ge
2$.  $CP$ violation is thus possible in the $\nu$MSM.  Third, the
out-of-equilibrium condition is provided here by the expansion of the
universe.  If the rate of sterile neutrino production $\Gamma_{\rm
sterile}$ is less than the rate of expansion of the universe $H$, then
sterile neutrinos do not interact enough to preserve equilibrium. 
This ratio can be estimated as follows:
\begin{table}
\caption{\label{Table:Parameters}}
\begin{ruledtabular}
\begin{tabular}{cccccc} A & B & $M_{IJ}$ & Yuk. + Mix. & Phases & Total \\ 
3 & 3 & 3 & 9 & 6 & 18 \\ 
3 & 2 & 2 & 6 & 3 & 11 \\ 
3 & 1 & 1 & 3 & 0 & 4 \\ 
2 & 2 & 2 & 4 & 2 & 8 \\ 
2 & 1 & 1 & 2 & 0 & 3 \\ 
1 & 2 & 2 & 2 & 1 & 5 \\ 
\end{tabular}
\end{ruledtabular}
\end{table}
\begin{eqnarray}
\label{eq:Out_of_equilibrium_criterion}
\frac{\Gamma_{\rm sterile}}{H} & \sim & \frac{f^{2}T}{T^2/M_{\rm Pl}}
 \;\;\sim\;\; 0.1,
\end{eqnarray}
where we used $T = T_{\rm sph} \sim 100$ GeV (temperature at which
sphalerons become inefficient) and $f = 10^{-9}$ (a possible Yukawa
coupling in the $\nu$MSM).  Since the ratio is inversely proportional
to the temperature, the sterile neutrinos are out-of-equilibrium for
all temperatures of interest.  Note that this ratio depends on $f$ and
that the Yukawa couplings in the $\nu$MSM are constrained by
observations \cite{Shaposhnikov_2008}.  The value used in the above
estimate is for a typical choice of parameters; there is a choice of
coupling  when sterile neutrinos equilibrate at $T > 100$ GeV. In the
latter case our formalism does not work and kinetic equations (or more sophisticated non-equilibrium quantum field theory tools) must be used.

\subsection{Outline of the Baryogenesis Scenario}
\label{Sec:Outline_scenario}
The scenario for baryon asymmetry generation in sterile neutrino
oscillations was proposed in Ref.~\cite{Akhmedov_etal_1998} and
developed in Refs~\cite{Asaka_Shaposhnikov_2005,Shaposhnikov_2008}. 
Just after reheating ($t_{i} = 0$), we have a thermal distribution of
Standard Model particles and we assume that there is no sterile
neutrino initially.  This is true if there is no source of sterile
neutrino during inflation, such as in recent models where the inflaton
is played by the Higgs \cite{Bezrukov_Shaposhnikov_2007,Bezrukov_etal_2008}.  There are
no experimental data that could constrain the initial conditions at
the moment.  The final result for the asymmetry of course depends on
the initial conditions.  Note that the formalism presented in
Sect.~\ref{Sec:Perturbative_formula} is valid for any initial distribution of sterile neutrinos.

During the following cosmological evolution, the three Sakharov
conditions are satisfied and a lepton asymmetry is produced via
coherent and resonant oscillations of active-sterile neutrinos.  The
necessary resonance condition for a sufficient asymmetry generation is
that two of the sterile neutrinos should be nearly degenerate in
mass.  This lepton asymmetry is converted into a baryon asymmetry via
sphaleron processes.  Since sphalerons become inefficient at a
temperature around $T_{\rm sph} \sim 100$ GeV, the conversion process
stops at around $t_{f} = t_{\rm sph} = 10^{13}$ GeV$^{-1}$.

This scenario is similar in spirit to ``usual'' or thermal
leptogenesis, although the physics is different.  In the case of
thermal leptogenesis, the parameters of the model are different (large
Yukawa couplings and large masses for the sterile neutrinos) and the
sterile neutrinos are initially in thermal equilibrium.  It is after
this initial period of thermal equilibrium that they go
out-of-equilibrium and decay, producing a lepton asymmetry.  This
period of out-of-equilibrium is relatively short.  In the case of the
$\nu$MSM, sterile neutrinos are not in equilibrium initially and
typically stay out-of-equilibrium (due to their very small couplings)
until sphaleron processes become inefficient.  The exact time at which
sterile neutrinos reach thermal equilibrium depends on the parameters
of the model (see Sect.~\ref{Sec:Validity_perturbation_theory}).

\section{Perturbative Formula for the (Lepton) Asymmetry}
\label{Sec:Perturbative_formula}

\subsection{Derivation of the Master Formula}
\label{Sec:Derivation_formula}
The total lepton asymmetry per unit volume is given by the difference
between the average number density of leptons minus the average number
density of anti-leptons. The asymmetry at time $t$ for a particular
flavor $\alpha$ of active neutrino $\nu_{\alpha}$ is:
\begin{eqnarray}
\label{eq:Asymmetry}
\Delta_{\alpha}(t,\vec{x}) & = & \mbox{Tr}\left[\hat{\rho}(t) 
\nu_{\alpha}^{\dagger}(t,\vec{x}) \nu_{\alpha}(t,\vec{x}) \right].
\end{eqnarray}
All fields in Eq.~(\ref{eq:Asymmetry}) are in the interaction
picture.  Here $\hat{\rho}(t)$ is the appropriate density operator.  The initial condition for the density matrix is expressed as:
\begin{eqnarray}
\label{eq:Initial_conditions}
\hat{\rho}(0) = \hat{\rho_{\rm S}}\otimes \hat{\rho}_{\rm SM},
\end{eqnarray}
where $\hat{\rho_{\rm S}}$ is the out-of-equilibrium density matrix
for the sterile neutrinos and $\hat{\rho}_{\rm SM} = e^{-\hat{H}_{\rm
SM}/T}$ is the usual equilibrium density operator for Standard Model
particles.  The precise form of $\hat{\rho_{\rm S}}$ is irrelevant for
the derivation of the master formula, but the final value of the
asymmetry is dependent on it.  For the case of the baryogenesis
scenario outlined in Sect.~\ref{Sec:Outline_scenario}, the density
matrix for sterile neutrinos is $\hat{\rho_{\rm S}}=|0\rangle\langle
0|$ at $t=0$, where $|0\rangle$ is a sterile neutrino vacuum state.  

Equations~(\ref{eq:Asymmetry}) and (\ref{eq:Initial_conditions})
constitutes the system that we want to solve (for any particular
flavor).  In equilibrium, the density operator is an exponential and
is time-independent; the machinery to solve such problems is very well
developed.  In the present case,
Eq.~(\ref{eq:Out_of_equilibrium_criterion}) shows that the system is
out-of-equilibrium, meaning that the density operator is
time-dependent and that usual equilibrium techniques fail.  A simple
way to see why such problems are hard is that propagators at finite
temperature depend on distribution functions.  The building blocks of
perturbation theory are propagators and vertices.  In
out-of-equilibrium situations, distribution functions are
time-dependent, implying that propagators change with time in a
non-trivial way.  Thus usual perturbation theory fails and resummation
techniques must be used
\cite{Calzetta_Hu_1988,Boyanovsky_deVega_2003,Berges_2004}, unless
changes in the propagators are small on timescales relevant for the problem at hand.  In the following, we treat
Eqs.~(\ref{eq:Asymmetry})-(\ref{eq:Initial_conditions}) perturbatively
and show in the next section under what condition perturbation theory
is valid.  

The first step is to find the time evolution of the density operator. 
It is given by the von Neumann equation:
\begin{eqnarray}
\label{eq:vonNeumann_equation}
i\frac{d\hat{\rho}(t)}{dt} & = & \left[\hat{H}_{I}(t),\hat{\rho}(t)\right],
\end{eqnarray}
where $\hat{H}(t) = \hat{H}_{0} + \hat{H}_{I}(t)$ is the interaction
Hamiltonian corresponding to the Lagrangian~(\ref{eq:Lagrangian}). 
All operators in Eq.~(\ref{eq:vonNeumann_equation}) are in the
interaction picture.  Note that we work in flat spacetime here; effects due to an expanding background can be incorporated using the usual procedure of expressing the equations in conformal time and going back to physical time at the end of the computation.  The rest of the derivation does not depend on the background.

We can find an iterative solution to
Eq.~(\ref{eq:vonNeumann_equation}):
\begin{eqnarray}
\label{eq:Iterative_solution}
\hat{\rho}(t) & = & \hat{\rho}(0) - i\int_{0}^{t} dt'\; \left[\hat{H}_{I}(t'),
\hat{\rho}(0)\right] - \int_{0}^{t} dt'\int_{0}^{t'}dt''\;
 \left[\hat{H}_{I}(t'),\left[\hat{H}_{I}(t''),
 \hat{\rho}(t'')\right]\right] + O(\hat{H}_{I}^{3}).
\end{eqnarray}
Note that the expansion parameter in Eq.~(\ref{eq:Iterative_solution})
is ``$\hat{H}_{I}t$''.  Thus the iterative
solution~(\ref{eq:Iterative_solution}) is also a perturbative solution
if the criterion ``$\hat{H}_{I}t \ll 1$'' is satisfied.  We use
quotation marks here to indicate that this is not a precise criterion
since $\hat{H}_{I}$ is an operator and $t$ is a number; we present a
more precise criterion in the next section.

Substituting the iterative solution~(\ref{eq:Iterative_solution}) back
into Eq.~(\ref{eq:Asymmetry}), we schematically get:
\begin{eqnarray}
\Delta_{\alpha}(t',\vec{x}) & = &  \mbox{Tr}\left[ \left(\hat{\rho}(0) 
+ \hat{\rho}^{(1)}(t') + \hat{\rho}^{(2)}(t') + ... \right) 
\nu_{\alpha}^{\dagger}(t',\vec{x})  \nu_{\alpha}(t',\vec{x})\right].
\end{eqnarray}
The first contribution is just an irrelevant infinite constant
independent of the temperature; it is an artefact of using the lepton
current to compute the asymmetry and we discard it in the following. 
The second contribution (and all contributions $\hat{\rho}^{(n)}$ with
$n$ odd) are automatically zero because they contain an odd number of
creation/annihilation operators.  The first non-trivial contribution
is thus (we only keep terms up to $O(\hat{H}_{I}^{2})$ in the
following):
\begin{eqnarray}
\label{eq:Asymmetry_1}
\Delta_{\alpha}(t',\vec{x}) & = &  \int_{0}^{t'}dt\int_{0}^{t}dt_{1}\; 
\mbox{Tr}\left[ \left(-[\hat{H}_{I}(t),[\hat{H}_{I}(t_{1}),\hat{\rho}(0)]] 
\right)\nu_{\alpha}^{\dagger}(t',\vec{x}) \nu_{\alpha}(t',\vec{x})\right].
\end{eqnarray}
To make progress, the interaction Hamiltonian need to be specified.  A
convenient form for our purposes is to decompose it as:
\begin{eqnarray}
\label{eq:Interaction_Hamiltonian}
\hat{H}_{I}(t) = \int d^{3}x\; \left(\bar{\nu}_{\beta}(t,\vec{x})
J_{\beta}(t,\vec{x}) + 
\bar{J}_{\beta}(t,\vec{x})\nu_{\beta}(t,\vec{x}) \right),
\end{eqnarray}
where $J$ contains all the information about the interaction except
the field associated to the desired asymmetry.  The exact form of $J$
is not important for the rest of the derivation.  Note that $J$
contains the out-of-equilibrium sterile neutrino field; in the case of
the $\nu$MSM, it also contains the Higgs field and the Yukawa
couplings.

Inserting this form for the interaction Hamiltonian into
Eq.~(\ref{eq:Asymmetry_1}) and using Wick's theorem, we obtain:
\begin{eqnarray}
\label{eq:Asymmetry_2}
\Delta_{\alpha}(t',\vec{x}) & = &  -\int_{0}^{t'}dt\int_{0}^{t}dt_{1}\int d^{3}y\int d^{3}y_{1}\; \nonumber \\
                            &   & \times 2\mbox{Re}\left[\langle J_{\beta}\bar{J}_{\gamma} \rangle \langle \nu_{\alpha}^{\dagger}\nu_{\gamma}\rangle \langle \nu_{\alpha }\bar{\nu}_{\beta}\rangle - \langle \bar{J}_{\beta}J_{\gamma} \rangle \langle \nu_{\alpha}^{\dagger}\nu_{\beta}\rangle \langle \nu_{\alpha}\bar{\nu}_{\gamma} \rangle \right. \nonumber \\
                            &   &  \left. \hspace{0.36in} - \langle \bar{J}_{\gamma}J_{\beta} \rangle \langle \nu_{\gamma}\nu_{\alpha}^{\dagger}\rangle \langle \nu_{\alpha}\bar{\nu}_{\beta} \rangle  + \langle J_{\gamma}\bar{J}_{\beta} \rangle \langle \nu_{\alpha}^{\dagger}\nu_{\beta}\rangle \langle \bar{\nu}_{\gamma}\nu_{\alpha} \rangle \right] + O(\hat{H}_{I}^{4}),
\end{eqnarray}
where we have used a condensed notation $J_{\beta} \equiv J_{\beta
a}(t,\vec{y})$ with the first and second indices being flavor and
spinor indices, respectively; $\alpha$ always refers to the external
flavor index and spacetime coordinates $(t',\vec{x})$, while
$\beta,\gamma,...$ correspond to internal flavor/spinor indices and to
spacetime coordinates $(t,\vec{y}),(t_{1},\vec{y}_{1}),...$.  The
equilibrium correlators $\langle \nu\bar{\nu} \rangle$ are the usual
active neutrino propagators.  The correlators $\langle J \bar{J}
\rangle$ contain the out-of-equilibrium sterile neutrino fields $N$
and potentially other fields $\Phi$ that are in equilibrium; they can
be decomposed as $\langle J \bar{J} \rangle =
\mbox{Tr}\left[\hat{\rho}_{\rm S}\otimes \hat{\rho}_{\rm SM} J
\bar{J}\right] \sim \mbox{Tr}\left[\hat{\rho}_{\rm S} N \bar{N}\right]
\otimes \mbox{Tr}\left[\hat{\rho}_{\rm SM} \Phi \bar{\Phi} \right]$. 
Note that for the scenario outlined in
Sect.~\ref{Sec:Outline_scenario} we have $\hat{\rho}_{\rm S} =
|0\rangle\langle 0|$ and thus zero temperature propagators can be used
to describe the evolution of the sterile neutrinos.  This replacement
is allowed if perturbation theory is valid; we come back to this point
in the next section.

It is possible to simplify Eq.~(\ref{eq:Asymmetry_2}) further by
making an assumption.  First note that for any propagator we have
$\langle \psi(x)\bar{\psi}(y) \rangle \sim \langle
\bar{\psi}(x)\psi(y) \rangle$, where the $\sim$ means everything is
equal except for their Dirac structures which are complex conjugate of
each other.  Looking at Eq.~(\ref{eq:Asymmetry_2}), we see that the
first and second terms and the third and fourth terms are the same
except that their coupling constants and their Dirac structures are
complex conjugate of each other.  In the following, we assume that the
Dirac structure computation for each term gives a real scalar; thus
for all practical purposes, the Dirac structure is irrelevant when
comparing these pairs of terms.  This assumption may be true in
general, but its validity must be checked explicitly for each
interaction Hamiltonian.

Taking into account the previous assumption and defining $J_{\beta}
\equiv F_{\beta I}\tilde{J}_{I}$, the asymmetry at second order in the
interaction Hamiltonian finally becomes:
\begin{eqnarray}
\label{eq:Asymmetry_H2}
\Delta_{\alpha}(t',\vec{x}) & = &  -\int_{0}^{t'}dt\int_{0}^{t}dt_{1}\int d^{3}y\int d^{3}y_{1}\; 2\mbox{Re}\left[2i\mbox{Im}(F_{\beta I}F_{\gamma J}^{*}) \langle \tilde{J}_{I}\bar{\tilde{J}}_{J} \rangle \langle \nu_{\alpha}^{\dagger}\nu_{\gamma}\rangle \langle \nu_{\alpha}\bar{\nu}_{\beta}\rangle \right. \nonumber \\
                            &   & \left. \hspace{2.2in} - 2i\mbox{Im}(F_{\gamma J}^{*}F_{\beta I}) \langle \bar{\tilde{J}}_{J}\tilde{J}_{I} \rangle \langle \nu_{\gamma}\nu_{\alpha}^{\dagger}\rangle \langle \nu_{\alpha}\bar{\nu}_{\beta} \rangle \right] \nonumber \\ 
                            & = & \int_{0}^{t'}dt\int_{0}^{t}dt_{1}\int d^{3}y\int d^{3}y_{1}\; 4\left[\mbox{Im}(F_{\beta I}F_{\gamma J}^{*}) \;\mbox{Im}(\langle \tilde{J}_{I}\bar{\tilde{J}}_{J} \rangle \langle \nu_{\alpha}^{\dagger}\nu_{\gamma}\rangle \langle \nu_{\alpha}\bar{\nu}_{\beta}\rangle) \right. \nonumber \\
                            &   & \left. \hspace{1.85in} - \mbox{Im}(F_{\gamma J}^{*}F_{\beta I}) \;\mbox{Im}(\langle \bar{\tilde{J}}_{J}\tilde{J}_{I} \rangle \langle \nu_{\gamma}\nu_{\alpha}^{\dagger}\rangle \langle \nu_{\alpha}\bar{\nu}_{\beta}) \rangle \right].
\end{eqnarray}
This last equation, or more specifically its $O(H_{I}^{4})$ version
(see the discussion below), are the main results of this paper.  It
expresses the lepton asymmetry for a particular flavor in terms of
Wigthman correlators.  If perturbation theory is valid, then these
correlators can be computed using conventional tools.  If the result
is zero, then higher order contributions in $\hat{H}_{I}$ must be
computed.

Formula~(\ref{eq:Asymmetry_H2}) is quite general, applicable in
principle to leptogenesis-type models.  The validity of perturbation
theory is the only assumption that enters into its derivation (in
addition to the minor assumption about the Dirac structure).

The last formula satisfies the third Sakharov condition.  As mentioned
previously, the $J$ operator contains the sterile neutrino operator,
and the sterile neutrinos are not in thermal equilibrium.  Assuming
that the sterile neutrinos are in equilibrium, then the $J$ operator
would obey the Kubo-Martin-Schwinger condition
\cite{Kubo_1957,Martin_Schwinger_1959} $\langle
J_{a}(t_{1})\bar{J}_{b}(t_{2}) \rangle = \langle
\bar{J}_{b}(t_{2})J_{a}(t_{1}+i\beta) \rangle$ ($\beta = 1/T$ is the
inverse temperature).  Using the Kubo-Martin-Schwinger condition and
time translation invariance, it can be shown that the asymmetry is
automatically zero in equilibrium.

Formula~(\ref{eq:Asymmetry_H2}) also neatly separates $CP$ violating
effects (contained in the imaginary part of the couplings) and
dynamical effects.  It is thus easy to see that it satisfies the
second Sakharov condition: if there is no $CP$ violating phase, then
the Yukawa couplings $F$'s are real and the formula gives
automatically zero.  Furthermore, using the fact that propagators are
diagonal in flavor space ($\langle \nu_{\alpha}\bar{\nu}_{\beta}
\rangle \propto \delta_{\alpha\beta}$ and $\langle
\tilde{J}_{I}\bar{\tilde{J}}_{J} \rangle \propto \delta_{IJ}$), we get
that $\mbox{Im}\left(F_{\beta I}F_{\beta I}^{*}\right) = 0$ and that
the formula for the asymmetry is automatically zero at
$O(H_{I}^{2})$.  This is another way of saying that a lepton asymmetry
is a quantum effect (generally coming from the interference of a tree
level diagram and a loop diagram).

The first non-trivial order in the expansion of the density operator
is thus $O(H_{I}^{4})$.  Expanding the iterative
solution~(\ref{eq:Iterative_solution}) up to $O(H_{I}^{4})$ and
repeating the same procedure, we finally obtain:
\begin{eqnarray}
\label{eq:Asymmetry_H4}
\Delta_{\alpha}(t',\vec{x}) & = &  -\int_{0}^{t'}dt\int_{0}^{t}dt_{1}\int_{0}^{t_{1}}dt_{2}\int_{0}^{t_{2}}dt_{3} \int d^{3}y\int d^{3}y_{1}\int d^{3}y_{2}\int d^{3}y_{3}\; \nonumber \\
                            &   & \times 4\left[\mbox{Im} (F_{\beta I}F_{\gamma J}F_{\delta K}^{*}F_{\epsilon L}^{*}) \;\mbox{Im}[\langle \tilde{J}_{I}\tilde{J}_{J}\bar{\tilde{J}}_{K}\bar{\tilde{J}}_{L} \rangle \right. \nonumber \\
                            &   & \hspace{0.35in} ( -\langle \nu_{\alpha}^{\dagger}\nu_{\delta}\rangle \langle\bar{\nu}_{\gamma}\nu_{\epsilon}\rangle + \langle \nu_{\alpha }^{\dagger}\nu_{\epsilon}\rangle \langle\bar{\nu}_{\gamma}\nu_{\delta}\rangle ) \left( \langle \bar{\nu}_{\beta}\nu_{\alpha}\rangle + \langle \nu_{\alpha}\bar{\nu}_{\beta}\rangle \right)] \nonumber \\
                            &   & \hspace{0.15in} + \mbox{Im} (F_{\beta I}F_{\gamma J}^{*}F_{\delta K}F_{\epsilon L}^{*}) \;\mbox{Im}[\langle \tilde{J}_{I}\bar{\tilde{J}}_{J}\tilde{J}_{K}\bar{\tilde{J}}_{L} \rangle \nonumber \\
                            &   & \hspace{0.35in} ( \langle \nu_{\alpha}^{\dagger}\nu_{\gamma}\rangle \langle\bar{\nu}_{\delta}\nu_{\epsilon}\rangle + \langle \nu_{\alpha }^{\dagger}\nu_{\epsilon}\rangle \langle\nu_{\gamma}\bar{\nu}_{\delta}\rangle ) \left( \langle \bar{\nu}_{\beta}\nu_{\alpha}\rangle + \langle \nu_{\alpha}\bar{\nu}_{\beta}\rangle \right)] \nonumber \\
                            &   & \hspace{0.15in} + \mbox{Im} (F_{\beta I}F_{\gamma J}^{*}F_{\delta K}^{*}F_{\epsilon L}) \;\mbox{Im}[\langle \tilde{J}_{I}\bar{\tilde{J}}_{J}\bar{\tilde{J}}_{K}\tilde{J}_{L} \rangle \nonumber \\
                            &   & \hspace{0.35in} ( \langle \nu_{\alpha}^{\dagger}\nu_{\gamma}\rangle \langle\nu_{\delta}\bar{\nu}_{\epsilon}\rangle - \langle \nu_{\alpha }^{\dagger}\nu_{\delta}\rangle \langle\nu_{\gamma}\bar{\nu}_{\epsilon}\rangle ) \left( \langle \bar{\nu}_{\beta}\nu_{\alpha}\rangle + \langle \nu_{\alpha}\bar{\nu}_{\beta}\rangle \right)] \nonumber \\
                            &   & \hspace{0.15in} + \mbox{Im} (F_{\epsilon L}^{*}F_{\beta I}F_{\gamma J}F_{\delta K}^{*}) \;\mbox{Im}[\langle \bar{\tilde{J}}_{L}\tilde{J}_{I}\tilde{J}_{J}\bar{\tilde{J}}_{K} \rangle \nonumber \\
                            &   & \hspace{0.35in} ( -\langle \nu_{\epsilon}\nu_{\alpha}^{\dagger}\rangle \langle\bar{\nu}_{\gamma}\nu_{\delta}\rangle + \langle \nu_{\alpha }^{\dagger}\nu_{\delta}\rangle \langle\nu_{\epsilon}\bar{\nu}_{\gamma}\rangle ) \left( \langle \bar{\nu}_{\beta}\nu_{\alpha}\rangle + \langle \nu_{\alpha}\bar{\nu}_{\beta}\rangle \right)] \nonumber \\
                            &   & \hspace{0.15in} + \mbox{Im} (F_{\epsilon L}^{*}F_{\beta I}F_{\gamma J}^{*}F_{\delta K}) \;\mbox{Im}[\langle \bar{\tilde{J}}_{L}\tilde{J}_{I}\bar{\tilde{J}}_{J}\tilde{J}_{K} \rangle \nonumber \\
                            &   & \hspace{0.35in} ( -\langle \nu_{\epsilon}\nu_{\alpha}^{\dagger}\rangle \langle\nu_{\gamma}\bar{\nu}_{\delta}\rangle - \langle \nu_{\alpha }^{\dagger}\nu_{\gamma}\rangle \langle\nu_{\epsilon}\bar{\nu}_{\delta}\rangle ) \left( \langle \bar{\nu}_{\beta}\nu_{\alpha}\rangle + \langle \nu_{\alpha}\bar{\nu}_{\beta}\rangle \right)] \nonumber \\
                            &   & \hspace{0.15in} + \mbox{Im} (F_{\epsilon L}F_{\beta I}F_{\gamma J}^{*}F_{\delta K}^{*}) \;\mbox{Im}[\langle \tilde{J}_{L}\tilde{J}_{I}\bar{\tilde{J}}_{J}\bar{\tilde{J}}_{K} \rangle \nonumber \\
                            &   & \hspace{0.35in} ( -\langle \nu_{\alpha}^{\dagger}\nu_{\gamma}\rangle \langle\bar{\nu}_{\epsilon}\nu_{\delta}\rangle + \langle \nu_{\alpha }^{\dagger}\nu_{\delta}\rangle \langle\bar{\nu}_{\epsilon}\nu_{\gamma}\rangle ) \left( \langle \bar{\nu}_{\beta}\nu_{\alpha}\rangle + \langle \nu_{\alpha}\bar{\nu}_{\beta}\rangle \right)] \nonumber \\
                            &   & \hspace{0.15in} + \mbox{Im} (F_{\delta K}^{*}F_{\beta I}F_{\gamma J}F_{\epsilon L}^{*}) \;\mbox{Im}[\langle \bar{\tilde{J}}_{K}\tilde{J}_{I}\tilde{J}_{J}\bar{\tilde{J}}_{L} \rangle \nonumber \\
                            &   & \hspace{0.35in} ( -\langle \nu_{\delta}\nu_{\alpha}^{\dagger}\rangle \langle\bar{\nu}_{\gamma}\nu_{\epsilon}\rangle + \langle \nu_{\alpha }^{\dagger}\nu_{\epsilon}\rangle \langle\nu_{\delta}\bar{\nu}_{\gamma}\rangle ) \left( \langle \bar{\nu}_{\beta}\nu_{\alpha}\rangle + \langle \nu_{\alpha}\bar{\nu}_{\beta}\rangle \right)] \nonumber \\
                            &   & \hspace{0.15in} + \mbox{Im} (F_{\delta K}^{*}F_{\beta I}F_{\gamma J}^{*}F_{\epsilon L}) \;\mbox{Im}[\langle \bar{\tilde{J}}_{K}\tilde{J}_{I}\bar{\tilde{J}}_{J}\tilde{J}_{L} \rangle \nonumber \\
                            &   & \hspace{0.35in} ( -\langle \nu_{\delta}\nu_{\alpha}^{\dagger}\rangle \langle\nu_{\gamma}\bar{\nu}_{\epsilon}\rangle - \langle \nu_{\alpha }^{\dagger}\nu_{\gamma}\rangle \langle\nu_{\delta}\bar{\nu}_{\epsilon}\rangle ) \left( \langle \bar{\nu}_{\beta}\nu_{\alpha}\rangle + \langle \nu_{\alpha}\bar{\nu}_{\beta}\rangle \right)] \nonumber \\
                            &   & \hspace{0.15in} + \mbox{Im} (F_{\delta K}F_{\beta I}F_{\gamma J}^{*}F_{\epsilon L}^{*}) \;\mbox{Im}[\langle \tilde{J}_{K}\tilde{J}_{I}\bar{\tilde{J}}_{J}\bar{\tilde{J}}_{L} \rangle \nonumber \\
                            &   & \hspace{0.35in} ( -\langle \nu_{\alpha}^{\dagger}\nu_{\gamma}\rangle \langle\bar{\nu}_{\delta}\nu_{\epsilon}\rangle + \langle \nu_{\alpha }^{\dagger}\nu_{\epsilon}\rangle \langle\bar{\nu}_{\delta}\nu_{\gamma}\rangle ) \left( \langle \bar{\nu}_{\beta}\nu_{\alpha}\rangle + \langle \nu_{\alpha}\bar{\nu}_{\beta}\rangle \right)] \nonumber \\
                            &   & \hspace{0.15in} + \mbox{Im} (F_{\epsilon L}^{*}F_{\delta K}^{*}F_{\beta I}F_{\gamma J}) \;\mbox{Im}[\langle \bar{\tilde{J}}_{L}\bar{\tilde{J}}_{K}\tilde{J}_{I}\tilde{J}_{J} \rangle \nonumber \\
                            &   & \hspace{0.35in} ( \langle \nu_{\delta}\nu_{\alpha}^{\dagger}\rangle \langle\nu_{\epsilon}\bar{\nu}_{\gamma}\rangle - \langle \nu_{\epsilon}\nu_{\alpha }^{\dagger}\rangle \langle\nu_{\delta}\bar{\nu}_{\gamma}\rangle ) \left( \langle \bar{\nu}_{\beta}\nu_{\alpha}\rangle + \langle \nu_{\alpha}\bar{\nu}_{\beta}\rangle \right)] \nonumber \\
                            &   & \hspace{0.15in} + \mbox{Im} (F_{\epsilon L}F_{\delta K}^{*}F_{\beta I}F_{\gamma J}^{*}) \;\mbox{Im}[\langle \tilde{J}_{L}\bar{\tilde{J}}_{K}\tilde{J}_{I}\bar{\tilde{J}}_{J} \rangle \nonumber \\
                            &   & \hspace{0.35in} ( \langle \nu_{\delta}\nu_{\alpha}^{\dagger}\rangle \langle\bar{\nu}_{\epsilon}\nu_{\gamma}\rangle + \langle \nu_{\alpha }^{\dagger}\nu_{\gamma}\rangle \langle\bar{\nu}_{\epsilon}\nu_{\delta}\rangle ) \left( \langle \bar{\nu}_{\beta}\nu_{\alpha}\rangle + \langle \nu_{\alpha}\bar{\nu}_{\beta}\rangle \right)] \nonumber \\
                            &   & \hspace{0.15in} + \mbox{Im} (F_{\epsilon L}^{*}F_{\delta K}F_{\beta I}F_{\gamma J}^{*}) \;\mbox{Im}[\langle \bar{\tilde{J}}_{L}\tilde{J}_{K}\tilde{J}_{I}\bar{\tilde{J}}_{J} \rangle \nonumber \\
                            &   & \left. \hspace{0.35in} ( \langle \nu_{\alpha}^{\dagger}\nu_{\gamma}\rangle \langle\nu_{\epsilon}\bar{\nu}_{\delta}\rangle - \langle \nu_{\epsilon}\nu_{\alpha }^{\dagger}\rangle \langle\bar{\nu}_{\delta}\nu_{\gamma}\rangle ) \left( \langle \bar{\nu}_{\beta}\nu_{\alpha}\rangle + \langle \nu_{\alpha}\bar{\nu}_{\beta}\rangle \right)] \right].
\end{eqnarray}
Formula~(\ref{eq:Asymmetry_H4}) allows the computation of the BAU up
to $O(f^{4})$ and is exact in all other couplings of the model.  Since
perturbation theory is assumed to be valid, it is possible to resolve
the 4-point functions $\langle
\tilde{J}_{I}\tilde{J}_{J}\bar{\tilde{J}}_{K}\bar{\tilde{J}}_{L}
\rangle$ into products of 2-point functions using Wick's theorem. 
Thus the only inputs in Eq.~(\ref{eq:Asymmetry_H4}) are the $F_{\alpha
I}$'s and the various propagators of the model.  The resulting terms
can be interpreted as Feynman diagrams.

The creation of the asymmetry takes place in the early universe where
a thermal bath of Standard Model particles is present.  It is well
known that resummed propagators (with thermal masses and dampings)
must be used in order to obtain correct results.  For instance,
interactions of particles with the medium may open new decay channels
that are otherwise kinematically forbidden (see for
example~\cite{Rychkov_Strumia_2007}).  The inclusion of damping is
also important to deal with so-called secular terms or ``pinch''
singularities (see Sect.~\ref{Sec:Validity_perturbation_theory}). 
Hard thermal loop resummations are moreover necessary to obtain gauge
invariant results in some cases~\cite{Braaten_Pisarski_1990}.

\begin{figure}
\scalebox{0.9}{\includegraphics{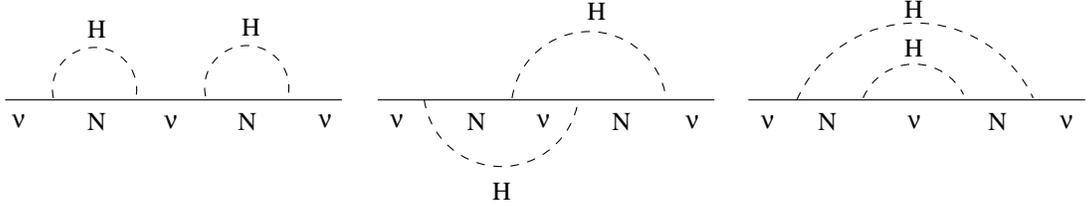}}
\caption{\label{fig:Diagrams_perturbative_formula} Typical Feynman diagrams appearing in the perturbative formula ~(\ref{eq:Asymmetry_H4}).  The solid lines correspond to active $\nu$ and sterile $N$ neutrino propagators and the dashed lines correspond to Higgs $H$ propagators.}
\end{figure}

\begin{figure}
\scalebox{0.9}{\includegraphics{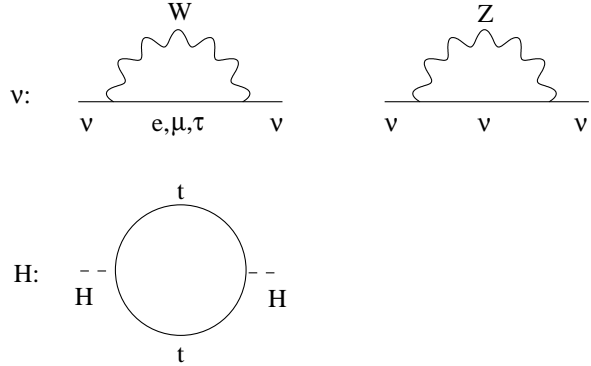}}
\caption{\label{fig:Diagrams_resummed} Self-energies that are resummed in the active neutrino $\nu$ and Higgs $H$ propagators.}
\end{figure}

To end this section, we discuss qualitatively the case of the $\nu$MSM
in order to illustrate the use of the perturbative
formula~(\ref{eq:Asymmetry_H4}) (for a similar discussion in a kinetic
theory setup, see \cite{Asaka_Shaposhnikov_2005,Shaposhnikov_2008}). 
We restrict ourselves to the symmetric phase here.  The typical
Feynman diagrams appearing in the perturbative formula are given in
Fig.~(\ref{fig:Diagrams_perturbative_formula}).  In order to compute
the asymmetry, we need the (Wightman) active/sterile neutrino and
Higgs propagators.  The exact form of these propagators depend on the
self-energies that are resummed into them.  For sterile neutrinos, no
resummation is necessary since the propagators are at zero temperature
and the self-energy corrections are small (i.e. $O(f^{2})$).  For
active neutrinos, it is necessary to include $W,Z$ boson and charged
lepton loops.  The absence of charged leptons would imply a new
symmetry for the $\nu$MSM, allowing the removal of $CP$ phases in
$F_{\alpha I}$ by rotating active neutrino fields and leading to a
vanishing asymmetry \footnote{In the case of 3 active and 3 sterile
neutrinos, we have 6 $CP$ violating phases (see
Table~\ref{Table:Parameters}).  In the absence of interactions between
charged leptons and $W$ bosons, the neutrino fields are not
constrained anymore and it is possible to make a unitary
transformation on them $\nu' = {\cal U}\nu$ (where ${\cal U}$ contains
3 real parameters and 6 phases).  This additional freedom in ${\cal
U}$ can be used to remove the 6 $CP$ phases in the Yukawa matrix
$F_{\alpha I}$ and make the asymmetry vanish.}.  For Higgs bosons, the
dominant contribution to the self-energy comes from top quark loops of
size $O(m_{t}^{2}T^{2}/v^{2})$; this is parametrically large compared
to $W,Z$ boson loops of size $O(m_{W,Z}^{4}/v^{2})$ and Higgs loops of
size $O(m_{H}^{4}/v^{2})$ (all the estimates are for large
temperatures).  The self-energies that need to be resummed in the
propagators are illustrated in Fig.~(\ref{fig:Diagrams_resummed}).

As can be seen from the above qualitative discussion, the computation
of the BAU in the $\nu$MSM is not a simple task.  For this reason and
as a warm up, we analyze a simpler toy model involving only Yukawa
interactions with sterile neutrinos and no other interaction.  The
full computation of the BAU in the $\nu$MSM will be the subject of a
separate publication.

\subsection{Validity of Perturbation Theory}
\label{Sec:Validity_perturbation_theory}

For the iterative solution~(\ref{eq:Iterative_solution}) to be also a
perturbative solution, the criterion ``$\hat{H}_{I}t \ll 1$'' needs to
be satisfied.  Said differently, it means that perturbation theory is
bound to break down for sufficiently long times; these are the
infamous secular terms that plague non-equilibrium quantum field
theory (e.g. \cite{Berges_2004}).  These secular terms also appear in
a different form in quantum field theory computations using the
real-time formalism of finite temperature field theory
\cite{Boyanovsky_etal_1999}.  In the real-time formalism (where time
is taken to go from minus infinity to plus infinity and back),
ill-defined products of delta functions (or ``pinch'' singularities)
arise naturally \cite{Altherr_Seibert_1994,Altherr_1995} and blow up
with the time-volume \cite{Greiner_Leupold_1998}.  Fortunately these
pinch singularities cancel in equilibrium due to the
Kubo-Martin-Schwinger condition \cite{Landsman_vanWeert_1987}.  Since
the Kubo-Martin-Schwinger condition is not valid out-of-equilibrium,
the pinch singularity problem remains for non-thermal systems and more
sophisticated methods must be used
\cite{Calzetta_Hu_1988,Boyanovsky_deVega_2003,Berges_2004}.

In the present paper, we use the von Neumann equation and time is
finite: if the final time is taken to be small, then secular terms
should also stay small (see \cite{Gelis_etal_2001} for a similar
argument phrased in terms of pinch singularities).  We can estimate
the size of the secular terms in the following way.  Roughly speaking,
``$\hat{H}_{I}t$'' gives the number of sterile neutrinos produced per
unit time ($\hat{H}_{I}$) times the time ($t$).  The total number of
sterile neutrinos produced is given by:
\begin{eqnarray}
n_{\rm sterile} & \sim & \int_{0}^{t_{\rm sph}} 
\Gamma_{\rm sterile}(t') dt' \;\;
\stackrel{\rm power\; laws}{\sim}\;\; 
\left.\frac{\Gamma_{\rm sterile}}{H}\right|_{t_{\rm sph}},
\end{eqnarray}
where in the last step we use the fact that both the rates of sterile
neutrino production and of the universe's expansion are power laws. 
Thus the size of secular terms (or equivalently the total number of
sterile neutrinos produced) is given by the out-of-equilibrium
criterion~(\ref{eq:Out_of_equilibrium_criterion}) evaluated at the
time when sphalerons become inefficient.  For $T = T_{\rm sph} = 100$
GeV and $f = 10^{-9}$, we get $n_{\rm sterile} \sim 0.1 \ll 1$ and
thus perturbation theory is justified in this part of parameter
space.  It also shows that the distribution function for the sterile
neutrinos does not evolve much during the relevant period for
baryogenesis and justifies our use of zero temperature propagators to
describe their evolution (in the case of the baryogenesis scenario
outlined in Sect.~ \ref{Sec:Outline_scenario}).

\section{Application to a Toy Model}
\label{Sec:Application}

Before using the formula for the asymmetry~(\ref{eq:Asymmetry_H4}) on
a realistic but more complicated model (such as the $\nu$MSM), we
would like to test it on a simpler (but unrealistic) model and verify
the range of validity of perturbation theory.  We thus make the
following simplifications to the model
Lagrangian~(\ref{eq:Lagrangian}).  First, we neglect interactions
(damping) with Standard Model particles.  This implies that the total
active lepton asymmetry is zero (see the discussion below
Eq.~(\ref{eq:Asymmetry_H4})) and only individual flavor asymmetries
are non-zero.  Second, we assume that the Higgs is non-dynamical,
making the Lagrangian quadratic in the fields.  This simplified model
thus describes a bunch of non-interacting harmonic oscillators and no
thermalization is possible.  Third, we reduce the number of free
parameters by reducing the number of active/sterile neutrinos. 
Looking at Table~(\ref{Table:Parameters}), the minimal model
containing only one $CP$ violating phase has one active and two
sterile neutrinos.  For reasons that will become clear in the next
section, the case with one active neutrino and two sterile neutrinos
trivially gives a vanishing asymmetry.  We thus opt for the
next-to-minimal model (i.e. two active and two sterile neutrinos) to
test our formula.  The Lagrangian for such a toy model is:
\begin{eqnarray}
\label{eq:Lagrangian_toy}
{\cal L}_{\rm toy} & = & \psi_{1R}^{\dagger}i\sigma^{\mu}\partial_{\mu}\psi_{1R} + \psi_{2R}^{\dagger}i\sigma^{\mu}\partial_{\mu}\psi_{2R} + \psi_{3L}^{\dagger}i\bar{\sigma}^{\mu}\partial_{\mu}\psi_{3L} + \psi_{4L}^{\dagger}i\bar{\sigma}^{\mu}\partial_{\mu}\psi_{4L} \nonumber \\
                   &   & - M\psi_{1R}^{T}i\sigma^{2}\psi_{2R} - \frac{\Delta M}{2}(\psi_{1R}^{T}i\sigma^{2}\psi_{1R} + \psi_{2R}^{T}i\sigma^{2}\psi_{2R}) \nonumber \\
                   &   & - fv\psi_{3L}^{\dagger}\psi_{1R} - \epsilon fv e^{i\eta}\psi_{3L}^{\dagger}\psi_{2R} - \delta fv \psi_{4L}^{\dagger}\psi_{2R} + h.c.,
\end{eqnarray}
where we take the Higgs field to be a constant $v$, $\psi_{1,2R}$ are
the sterile neutrino right-handed Weyl fields, $\psi_{3,4L}$ are the
active neutrino left-handed Weyl fields,  $M$ is the common mass of
the sterile neutrinos, $\Delta M$ is the sterile neutrino mass
difference, $f$ and $\epsilon$ and $\delta$ are Yukawa couplings and
$\eta$ is a $CP$ violating phase.

This toy model has five real parameters and one $CP$ violating phase
\footnote{Note that the toy model has 5+1 parameters instead of the
6+2 parameters indicated in Table~(\ref{Table:Parameters}).  This
reduction in the number of parameters is obtained by taking linear
combinations of the neutrino fields and rearranging the Lagrangian. 
It is only possible in the absence of damping.} (compare this to the 8
real parameters and 3 phases of the $\nu$MSM including constraints
from dark matter abundance).  The physics of leptogenesis in this toy
model is thus simpler.  For instance, if $\eta$ or $\epsilon$ or
$\Delta M$ are sent to zero, then the Lagrangian does not contain any
$CP$ or baryon number violating terms anymore and the asymmetry
vanishes (if $\delta$ is zero then we come back to the one active
neutrino case and the asymmetry is also zero).  These features should
also appear in the solution.  

We can solve this model in two ways.  Since the Higgs takes its
expectation value, the Lagrangian is quadratic in the fields and it is
possible to solve the system exactly.  If the Yukawa coupling $f$ is
small, then the system can also be solved perturbatively using the
master formula for the lepton asymmetry presented in
Sect.~\ref{Sec:Derivation_formula}.  The exact and perturbative
results can be compared and should match in some time interval.  This
is what we present in the next sections.

\subsection{Exact Solution}
\label{Sec:Exact_solution}
As in the perturbative case, the lepton asymmetry (for the active
flavor $a=3,4$) is given by the average lepton current density:
\begin{eqnarray}
\label{eq:Asymmetry_exact}
\Delta_{\alpha}(t,\vec{x}) & = & \mbox{Tr}\left[\rho(0) 
\psi_{\alpha L}^{\dagger}(t,\vec{x})\psi_{\alpha L}(t,\vec{x})\right],
\end{eqnarray}
where all fields are in the Heisenberg picture.  The calculation of
the asymmetry is now purely quantum mechanical and very similar to
neutrino oscillation computations (e.g. \cite{Giunti_Kim_2007}).  For
times $t < 0$, there is no interaction and there is no sterile
neutrino.  At $t = 0$, the interaction is adiabatically switched on
and sterile neutrino fields are initially in flavor eigenstates (in
which the thermal averages are defined).  This last point is very
important because otherwise the problem is trivial.  Indeed, since the
Lagrangian~(\ref{eq:Lagrangian_toy}) can be diagonalized such that all
four masses are real (see below), then it means that all $CP$
violating phases can be absorbed and there is no asymmetry if the
thermal averages are defined in the mass basis.  In the case where the
thermal averages are defined in the initial flavor basis, the
asymmetry is in principle non-zero.

The goal is to compute the asymmetry at some time $t > 0$, hence the
fields $\eta_{3L}(t,\vec{x})$ must be time evolved from $0$ to some
time $t$.  Time evolution in quantum mechanics is most conveniently
done using energy eigenstates.  In order to obtain the energy
eigenstates, we need the exact masses.  The mass matrix ${\cal M}$
in~(\ref{eq:Lagrangian_toy}) is:
\begin{eqnarray}
\label{eq:Mass_matrix}
{\cal M} & = & \left(\begin{array}{cccc} 
\Delta M & M & fv & 0 \\ 
M & \Delta M & \epsilon fv e^{-i\eta} & \delta fv \\ 
fv & \epsilon fv e^{-i\eta} & 0 & 0 \\ 
0 & \delta fv & 0 & 0 \end{array}\right).
\end{eqnarray}
The above symmetric complex mass matrix can be diagonalized using
Takagi's factorization ${\cal M} = {\cal U} {\cal D} {\cal U}^{T}$
where ${\cal U}$ is unitary (e.g. \cite{Cahill_2000}).  The results
for the masses are:
\begin{eqnarray}
\label{eq:Exact_masses}
\tilde{m}_{1} & = & M + \frac{f^2v^2}{2M}(1+\epsilon^2+\delta^2+2\rho^2), 
\nonumber \\
\tilde{m}_{2} & = & M + \frac{f^2v^2}{2M}(1+\epsilon^2+\delta^2-2\rho^2), 
\nonumber \\
\tilde{m}_{3} & = & \frac{f^2v^2}{M}(\epsilon + \sqrt{\epsilon^2+\delta^2}), 
\nonumber \\
\tilde{m}_{4} & = & \tilde{m}_{3}\left(\frac{\delta^2}
{2\epsilon^2+\delta^2+2\epsilon\sqrt{\epsilon^2+\delta^2}} \right),
\end{eqnarray}
where we used the parametrization $\Delta M \equiv
\left(\frac{f^{2}v^{2}(\epsilon+\delta)}{M}\right)\kappa$ and defined
$\rho e^{i\theta} \equiv \sqrt{\epsilon e^{i\eta} +
(\epsilon+\delta)\kappa}$.  Flavor and energy eigenstates are related
as $\psi_{R} = {\cal U}\tilde{\psi}_{R}$ and $\psi_{L} = {\cal
U}^{*}\tilde{\psi}_{L}$ where $\psi_{R} \equiv (\psi_{1R} \;\;
\psi_{2R} \;\; i\sigma^{2}\psi_{3L}^{*} \;\;
i\sigma^{2}\psi_{4L}^{*})^{T}$ and $\psi_{L} \equiv
(-i\sigma^{2}\psi_{1R}^{*} \;\; -i\sigma^{2}\psi_{2R}^{*} \;\;
\psi_{3L} \;\; \psi_{3L})^{T}$.  In the basis where the mass matrix is
diagonal, the neutrino fields are given by:
\begin{equation}
\label{eq:Neutrino_field_diagonal}
\tilde{\psi}_{iL}(t,\vec{x}) =  \int\frac{d^{3}p}{(2\pi)^{3}}
\frac{1}{\sqrt{2\tilde{E}_{p}^{i}}} \sum_{h\pm 1}
\left[\tilde{a}_{p,h,i}\sqrt{\tilde{E}_{p}^{i}-|\vec{p}|h}
\chi_{(h)}(\vec{p})e^{-ip\cdot x} + 
h\tilde{a}_{p,h,i}^{\dagger}\sqrt{\tilde{E}_{p}^{i}
+|\vec{p}|h}\chi_{(-h)}(\vec{p})e^{ip\cdot x} \right],
\end{equation}
where $h$ is the helicity and the $\chi_{(h)}(\vec{p})$'s are
two-component helicity eigenstate spinors.  The exact energies are
given by $\tilde{E}_{p}^{i} = \sqrt{|\vec{p}|^{2} +
\tilde{m}_{i}^{2}}$ and the exact masses $\tilde{m}_{i}$'s are given
in Eq.~(\ref{eq:Exact_masses}).   

After $t = 0$, the initial flavor eigenstates become linear
combinations of energy eigenstates and start to oscillate, with each
energy eigenstate oscillating with a different frequency depending on
the exact masses.  At time $t$, we have:
\begin{eqnarray}
\label{eq:Asymmetry_exact_1}
\Delta_{\alpha}(t,\vec{x}) & = & \mbox{Tr}\left[\rho(0) 
[\tilde{\psi}_{L}^{\dagger}(t,\vec{x}){\cal U}^{T}]_{\alpha}
[{\cal U}^{*}\tilde{\psi}_{L}(t,\vec{x})]_{\alpha}\right].
\end{eqnarray}
In the last expression, the neutrino fields are expressed in terms of
creation/annihilation operators $\tilde{a}_{p,h,i}$ that
create/annihilate quanta corresponding to states with a certain mass. 
To do the thermal averages, it is necessary to re-express these
``mass'' creation/annihilation operators as linear combinations of
``flavor'' creation/annihilation operators $a_{p,h,i}$.  This is done
using Bogoliubov-type transformations.  The appropriate
transformations are:
\begin{eqnarray}
\label{eq:Bogoliubov_transformations}
\tilde{a}_{p,h,i} & = & \frac{\sqrt{\tilde{E}_{i}+|\vec{p}|h}}{\sqrt{2\tilde{E}_{i}}} \left[\sum_{j=1,2} ({\cal U}^{\dagger}{\cal V})_{ij}\frac{1}{\sqrt{2E_{j}}} \left(a_{p,h,j}\sqrt{E_{j}+|\vec{p}|h} - h\phi(\vec{p},h)a_{-p,h,j}^{\dagger}\sqrt{E_{j}-|\vec{p}|h}\right) \right. \nonumber \\
                  &   & \hspace{0.75in} \left. + \sum_{k=3,4} ({\cal U}^{\dagger}{\cal V})_{ik} \left(b_{p,1,k}\delta_{1,h} + \phi(\vec{p},-1)a_{-p,-1,k}^{\dagger}\delta_{-1,h}\right)\right] \nonumber \\
                  &  & + \frac{\sqrt{\tilde{E}_{i}-|\vec{p}|h}}{\sqrt{2\tilde{E}_{i}}} \left[\sum_{j=1,2} ({\cal U}^{T}{\cal V^{*}})_{ij}\frac{1}{\sqrt{2E_{j}}} \left(a_{p,h,j}\sqrt{E_{j}-|\vec{p}|h} + h\phi(\vec{p},h)a_{-p,h,j}^{\dagger}\sqrt{E_{j}+|\vec{p}|h}\right) \right. \nonumber \\
                  &   & \hspace{0.75in} \left. + \sum_{k=3,4} ({\cal U}^{T}{\cal V^{*}})_{ik} \left(a_{p,-1,k}\delta_{-1,h} + \phi(\vec{p},1)b_{-p,1,k}^{\dagger}\delta_{1,h}\right)\right],
\end{eqnarray}
where ${\cal V}$ is the matrix that diagonalizes the free part (i.e.
$f=0$) of the Lagrangian ($\ref{eq:Lagrangian_toy}$), $E_{i}=
\sqrt{|\vec{p}|^{2} + m_{i}^{2}}$ are the eigenenergies corresponding
to the masses of the free Lagrangian and $\phi(\vec{p},h)$ is some
phase that satisfies $\phi(-\vec{p},-h)=\phi^{*}(\vec{p},h)$ and
$\phi(\vec{p},-h)=-\phi^{*}(\vec{p},h)$ \cite{Giunti_Kim_2007}.  One
can verify that these transformations preserve the canonical
anti-commutation relations.

Inserting the neutrino field
operator~(\ref{eq:Neutrino_field_diagonal}) into
Eq.~(\ref{eq:Asymmetry_exact_1}) and using the relations between the
two sets of creation/annihilation
operators~(\ref{eq:Bogoliubov_transformations}), the asymmetry becomes
(after some algebra):
\begin{eqnarray}
\label{eq:Asymmetry_exact_final}
\Delta_{\alpha}(t,\vec{x}) & = & -\sum_{i<j=1}^{2+A} \sum_{\kappa=3}^{2+A} \int\frac{d^{3}p}{(2\pi)^{3}} \left[\frac{2|\vec{p}|}{\tilde{E}_{i}\tilde{E}_{j}} n(E_{\kappa}) \mbox{Im}\left({\cal U}_{\alpha i}{\cal U}_{\alpha j}^{*}{\cal U}_{\kappa i}^{*}{\cal U}_{\kappa j}\right)   \right.  \nonumber \\
              &   &  \hspace{1.4in} \left. \left((\tilde{E}_{i}+\tilde{E}_{j})\sin{(\tilde{E}_{i}-\tilde{E}_{j})t}  -(\tilde{E}_{i}-\tilde{E}_{j})\sin{(\tilde{E}_{i}+\tilde{E}_{j})t} \right) \right].
\end{eqnarray}
This is the final result for the exact lepton asymmetry.  The above
formula is true for two sterile neutrinos and any number of active
neutrino flavors $A$.  The solution contains a $CP$ structure part and
a dynamical part consisting of two oscillating functions (one with a
large amplitude and small frequency and one with a small amplitude and
large frequency) for each value of $i,j$.  We note that the $CP$
structure part $\mbox{Im}\left({\cal U}_{\alpha i}{\cal U}_{\alpha
j}^{*}{\cal U}_{\kappa i}^{*}{\cal U}_{\kappa j}\right)$ is similar to
the one obtained in Ref.~\cite{Asaka_Shaposhnikov_2005}.  We also
immediately see that the total active lepton number is zero, i.e.
$\sum_{\alpha}\Delta_{\alpha}(t) \propto \sum_{\alpha}
\sum_{i<j=1}^{2+A} \sum_{\kappa =3}^{2+A} \mbox{Im}\left({\cal
U}_{\alpha i}{\cal U}_{\alpha j}^{*}{\cal U}_{\kappa i}^{*}{\cal
U}_{\kappa j}\right) = 0$.  In particular, we get $\Delta_{\alpha}(t)
= 0$ for only one flavor of active neutrinos; this explain our study
of the two active and two sterile neutrino case.

\subsection{Perturbative Solution}
\label{Sec:Perturbative_solution}

If the Yukawa coupling $f$ is small, then the active-sterile neutrino
interactions are small and can be treated as a perturbation. 
Diagonalizing the quadratic part of the
Lagrangian~(\ref{eq:Lagrangian_toy}) with the transformations
$\psi_{1R} \rightarrow (\eta_{1R}+i\eta_{2R})/\sqrt{2}$ and $\psi_{2R}
\rightarrow (\eta_{1R}-i\eta_{2R})/\sqrt{2}$ (the active neutrino
fields stay the same $\psi_{3,4L} \rightarrow \eta_{3,4L}$), we
obtain:
\begin{eqnarray}
{\cal L}_{\rm toy} & = &  \eta_{1R}^{\dagger}i\sigma^{\mu}\partial_{\mu}\eta_{1R} + \eta_{2R}^{\dagger}i\sigma^{\mu}\partial_{\mu}\eta_{2R} + \eta_{3L}^{\dagger}i\bar{\sigma}^{\mu}\partial_{\mu}\eta_{3L} + \eta_{4L}^{\dagger}i\bar{\sigma}^{\mu}\partial_{\mu}\eta_{4L} \nonumber \\
                   &   & - \frac{(M+\Delta M)}{2}\eta_{1R}^{T}i\sigma^{2}\eta_{1R} - \frac{(M-\Delta M)}{2}\eta_{2R}^{T}i\sigma^{2}\eta_{2R} \nonumber \\
                   &   & - \frac{fv}{\sqrt{2}}\eta_{3L}^{\dagger}(\eta_{1R}+i\eta_{2R}) - \frac{\epsilon fv e^{i\eta}}{\sqrt{2}}\eta_{3L}^{\dagger}(\eta_{1R}-i\eta_{2R}) - \frac{\delta fv}{\sqrt{2}}\eta_{4L}^{\dagger}(\eta_{1R}-i\eta_{2R}) + h.c.,
\end{eqnarray}
The perturbative masses are $m_{1,2} = M \pm \Delta M$ for the sterile
neutrinos and zero for the active neutrinos.  The interaction
Hamiltonian can be obtained from the above Lagrangian:
\begin{eqnarray}
\label{eq:Interaction_Hamiltonian_1}
H_{I} & = & \eta_{3L}^{\dagger}J_{3} + J_{3}^{\dagger}\eta_{3L} 
+ \eta_{4L}^{\dagger}J_{4} + J_{4}^{\dagger}\eta_{4L},
\end{eqnarray}
where the $J$ operators are given by:
\begin{eqnarray}
\label{eq:J_operator}
J_{3} & = & \frac{fv}{\sqrt{2}}(1+\epsilon e^{i\eta})\eta_{1R} + \frac{ifv}{\sqrt{2}}(1-\epsilon e^{i\eta})\eta_{2R} \;\;\equiv \;\; F_{31}\eta_{1R} + F_{32}\eta_{2R},  \nonumber \\
J_{4} & = & \frac{\delta fv}{\sqrt{2}}\eta_{1R} - \frac{i\delta fv}{\sqrt{2}}\eta_{2R} \;\;\equiv \;\; F_{41}\eta_{1R} + F_{42}\eta_{2R}.
\end{eqnarray}
Substituting the operators~(\ref{eq:J_operator}) in
Eq.~(\ref{eq:Asymmetry_H4}), it is a straightforward but tedious
exercice to compute the asymmetry.  In the following we present the
computation of (a part of) the first term in
Eq.~(\ref{eq:Asymmetry_H4}); the others are done in a similar way. 
Note that the assumption about the Dirac structure is verified here
(see the discussion following Eq.~(\ref{eq:Asymmetry_2})).  The first
term of Eq.~(\ref{eq:Asymmetry_H4}) is:
\begin{eqnarray}
\label{eq:Asymmetry_perturbative_1}
\Delta_{\alpha}(t',\vec{x})_{1} & = & -\int_{0}^{t'}dt \int_{0}^{t}dt_{1} \int_{0}^{t_{1}}dt_{2} \int_{0}^{t_{2}}dt_{3} \int d^{3}y \int d^{3}y_{1} \int d^{3}y_{2} \int d^{3}y_{3} \;  \nonumber \\
                                &   & \left[-4\mbox{Im}\left(F_{\alpha I}F_{\gamma J}F_{\alpha K}^{*}F_{\gamma N}^{*}\right) \; \mbox{Im} \left(\langle 0|\tilde{J}_{aI}(y)\tilde{J}_{bJ}(y_{1})\tilde{J}_{cK}^{\dagger}(y_{2})\tilde{J}_{dN}^{\dagger}(y_{3})|0\rangle \right. \right. \nonumber \\
                                &   & \left. \langle\eta_{\alpha L e }^{\dagger}(x)\eta_{\alpha  L c}(y_{2})\rangle \langle\eta_{\gamma L b}^{\dagger}(y_{1})\eta_{\gamma L d}(y_{3})\rangle (\langle \eta_{\alpha L a}^{\dagger}(y)\eta_{\alpha L e}(x)\rangle + \langle \eta_{\alpha L e}(x)\eta_{\alpha L a}^{\dagger}(y)\rangle) \right) \nonumber \\
                                &   & + 4\mbox{Im}\left(F_{\alpha I}F_{\gamma J}F_{\gamma K}^{*}F_{\alpha N}^{*}\right) \; \mbox{Im} \left(\langle 0|\tilde{J}_{aI}(y)\tilde{J}_{bJ}(y_{1})\tilde{J}_{cK}^{\dagger}(y_{2})\tilde{J}_{dN}^{\dagger}(y_{3})|0\rangle \right. \nonumber \\
                                &   & \left. \left. \langle\eta_{\alpha L e }^{\dagger}(x)\eta_{\alpha  L d}(y_{3})\rangle \langle\eta_{\gamma L b}^{\dagger}(y_{1})\eta_{\gamma L c}(y_{2})\rangle (\langle \eta_{\alpha L a}^{\dagger}(y)\eta_{\alpha L e}(x)\rangle + \langle \eta_{\alpha L e}(x)\eta_{\alpha L a}^{\dagger}(y)\rangle) \right) \right], \nonumber \\
\end{eqnarray}
where $a,b,...,e$ are spinor indices, $\alpha,\gamma$ refer to active neutrino flavors and $I,J,K,N$ refer to sterile neutrino flavors.  The case where $\alpha = \gamma$ corresponds to one active neutrino flavor only and gives zero.  For definiteness we thus consider $\alpha = 3$ and $\gamma = 4$ in the following.  Keeping only non-zero contributions coming from the imaginary part of the coupling constants, we obtain:
\begin{eqnarray}
\label{eq:Asymmetry_perturbative_2}
\Delta_{\alpha}(t',\vec{x})_{1} & = & -\int_{0}^{t'}dt \int_{0}^{t}dt_{1} \int_{0}^{t_{1}}dt_{2} \int_{0}^{t_{2}}dt_{3} \int d^{3}y \int d^{3}y_{1} \int d^{3}y_{2} \int d^{3}y_{3} \;  \nonumber \\
                                &   & \left[-4\mbox{Im}\left(F_{31}F_{41}F_{32}^{*}F_{42}^{*}\right) \; \mbox{Im} \left(\langle 0|\eta_{1Ra}(y)\eta_{1Rb}(y_{1})\eta_{2Rc}^{\dagger}(y_{2})\eta_{2Rd}^{\dagger}(y_{3})|0\rangle \right. \right. \nonumber \\
                                &   & \left. \langle\eta_{3L e}^{\dagger}(x)\eta_{3L c}(y_{2})\rangle \langle\eta_{4L b}^{\dagger}(y_{1})\eta_{4L d}(y_{3})\rangle (\langle \eta_{3L a}^{\dagger}(y)\eta_{3L e}(x)\rangle + \langle \eta_{3L e}(x)\eta_{3L a}^{\dagger}(y)\rangle) \right) \nonumber \\
                                &   & -4\mbox{Im}\left(F_{31}F_{42}F_{32}^{*}F_{41}^{*}\right) \; \mbox{Im} \left(\langle 0|\eta_{1Ra}(y)\eta_{2Rb}(y_{1})\eta_{2Rc}^{\dagger}(y_{2})\eta_{1Rd}^{\dagger}(y_{3})|0\rangle \right. \nonumber \\
                                &   & \left. \langle\eta_{3L e}^{\dagger}(x)\eta_{3L c}(y_{2})\rangle \langle\eta_{4L b}^{\dagger}(y_{1})\eta_{4L d}(y_{3})\rangle (\langle \eta_{3L a}^{\dagger}(y)\eta_{3L e}(x)\rangle + \langle \eta_{3L e}(x)\eta_{3L a}^{\dagger}(y)\rangle) \right) \nonumber \\
                                &   & +4\mbox{Im}\left(F_{31}F_{41}F_{32}^{*}F_{42}^{*}\right) \; \mbox{Im} \left(\langle 0|\eta_{1Ra}(y)\eta_{1Rb}(y_{1})\eta_{2Rc}^{\dagger}(y_{2})\eta_{2Rd}^{\dagger}(y_{3})|0\rangle \right.  \nonumber \\
                                &   & \left. \langle\eta_{3L e}^{\dagger}(x)\eta_{3L d}(y_{3})\rangle \langle\eta_{4L b}^{\dagger}(y_{1})\eta_{4L c}(y_{2})\rangle (\langle \eta_{3L a}^{\dagger}(y)\eta_{3L e}(x)\rangle + \langle \eta_{3L e}(x)\eta_{3L a}^{\dagger}(y)\rangle) \right) \nonumber \\
                                &   & +4\mbox{Im}\left(F_{31}F_{42}F_{32}^{*}F_{41}^{*}\right) \; \mbox{Im} \left(\langle 0|\eta_{1Ra}(y)\eta_{2Rb}(y_{1})\eta_{2Rc}^{\dagger}(y_{2})\eta_{1Rd}^{\dagger}(y_{3})|0\rangle \right. \nonumber \\
                                &   & \left. \langle\eta_{3L e}^{\dagger}(x)\eta_{3L d}(y_{3})\rangle \langle\eta_{4L b}^{\dagger}(y_{1})\eta_{4L c}(y_{4})\rangle (\langle \eta_{3L a}^{\dagger}(y)\eta_{3L e}(x)\rangle + \langle \eta_{3L e}(x)\eta_{3L a}^{\dagger}(y)\rangle) \right) \nonumber \\
                                &   & \left. + \mbox{(same terms with $1R\leftrightarrow 2R$ and $F_{\alpha 1}\leftrightarrow F_{\alpha 2}$)} \frac{}{}\right].
\end{eqnarray}
Assuming the validity of perturbation theory, all the neutrino fields
are free fields and we can use Wick's theorem to decompose the above
4-point functions into products of 2-point functions.  Since the
system is translationally invariant, we can do the Fourier transform
over space.  Concentrating on the first of the eight terms in
Eq.~(\ref{eq:Asymmetry_perturbative_2}), we have:
\begin{eqnarray}
\label{eq:Asymmetry_perturbative_3}
\Delta_{\alpha}(t',\vec{x})_{1,1} & = & \int_{0}^{t'}dt \int_{0}^{t}dt_{1} \int_{0}^{t_{1}}dt_{2} \int_{0}^{t_{2}}dt_{3} \int \frac{d^{3}p}{(2\pi)^{3}} \; 4\mbox{Im}\left(F_{31}F_{41}F_{32}^{*}F_{42}^{*}\right) \nonumber \\
                                &   & \mbox{Im} \left[\langle\eta_{3L e}^{\dagger}\eta_{3L c}\rangle(\vec{p})\; \langle 0|\eta_{2Rc}^{\dagger}\eta_{2Rd}^{\dagger}|0\rangle(\vec{p})\; \langle\eta_{4L b}^{\dagger}\eta_{4L d}\rangle(-\vec{p})\;  \right. \nonumber \\
                                &   & \left. \langle 0|\eta_{1Ra}\eta_{1Rb}|0\rangle(-\vec{p})(\langle \eta_{3L a}^{\dagger}\eta_{3L e}\rangle(\vec{p}) + \langle \eta_{3L e}\eta_{3L a}^{\dagger}\rangle)(\vec{p}) \right].
\end{eqnarray}
To make progress, we need the form of the propagators for massless
left-handed active neutrinos and for massive right-handed sterile
neutrinos.  These are obtained from the expansion of the fields in
terms of creation/annihilation operators (cf.
Eq.~(\ref{eq:Neutrino_field_diagonal})).  The propagators are:
\begin{eqnarray}
\label{eq:Propagators}
\langle \eta_{L}(t_{1})\eta_{L}^{\dagger}(t_{2})\rangle(\vec{p}) & = & (1-n(E_{p}))\chi_{(-1)}(\vec{p})\chi_{(-1)}^{\dagger}(\vec{p})e^{-iE_{p}(t_{1}-t_{2})} \nonumber \\ 
                                                                 &   & + n(E_{p})\chi_{(-1)}(-\vec{p})\chi_{(-1)}^{\dagger}(-\vec{p})e^{iE_{p}(t_{1}-t_{2})}, \nonumber \\
\langle \eta_{R}(t_{1})\eta_{R}(t_{2})\rangle(\vec{p}) & = & -\frac{1}{2E_{p}} \sum_{h}\left[(1-n(E_{p}))hm \chi_{(h)}(\vec{p})\chi_{(-h)}(\vec{p})e^{-iE_{p}(t_{1}-t_{2})} \right. \nonumber \\
                                                       &   & \hspace{0.5in} \left. + n(E_{p})hM\chi_{(-h)}(-\vec{p})\chi_{(h)}(-\vec{p})e^{iE_{p}(t_{1}-t_{2})}  \right], \nonumber \\
\langle \eta_{R}^{\dagger}(t_{1})\eta_{R}^{\dagger}(t_{2})\rangle(\vec{p}) & = & -\frac{1}{2E_{p}} \sum_{h}\left[(1-n(E_{p}))hm \chi_{(-h)}^{\dagger}(\vec{p})\chi_{(h)}^{\dagger}(\vec{p})e^{-iE_{p}(t_{1}-t_{2})} \right. \nonumber \\
                                                       &   & \hspace{0.5in} \left. + n(E_{p})hM\chi_{(h)}^{\dagger}(-\vec{p})\chi_{(-h)}^{\dagger}(-\vec{p})e^{iE_{p}(t_{1}-t_{2})}  \right].
\end{eqnarray}
Inserting these propagators in
Eq.~(\ref{eq:Asymmetry_perturbative_3}), we obtain:
\begin{eqnarray}
\label{eq:Asymmetry_perturbative_4}
\Delta(t',\vec{x})_{1,1} & = & -\int_{0}^{t'}dt \int_{0}^{t}dt_{1} \int_{0}^{t_{1}}dt_{2} \int_{0}^{t_{2}}dt_{3} \int \frac{d^{3}p}{(2\pi)^{3}} \; 4\mbox{Im}\left(F_{31}F_{41}F_{32}^{*}F_{42}^{*}\right)  \nonumber \\
                                 &   & \mbox{Im}\left[\frac{m_{1}m_{2}}{4E_{1}E_{2}}\left( (1-n(E_{3}))(1-n(E_{4})) e^{i(-E_{1}-E_{3})t}e^{i(E_{1}-E_{4})t_{1}}e^{i(-E_{2}+E_{3})t_{2}}e^{i(E_{2}+E_{4})t_{3}} \right. \right. \nonumber \\
                                 &   & \left. \left.  \hspace{0.8in} + n(E_{3})n(E_{4}) e^{i(-E_{1}+E_{3})t}e^{i(E_{1}+E_{4})t_{1}}e^{i(-E_{2}-E_{3})t_{2}}e^{i(E_{2}-E_{4})t_{3}} \right)\right].
\end{eqnarray}
The integrals over time are trivially done; the result is:
\begin{eqnarray}
\label{eq:Asymmetry_perturbative_5}
\Delta(t',\vec{x})_{1,1} & = & -2f^{4}v^{4} \epsilon\delta^{2}\sin\eta \int\frac{d^{3}p}{(2\pi)^{3}}\; \frac{m_{1}m_{2}}{4E_{1}E_{2}} \left[(1-n(E_{3}))(1-n(E_{4})) \frac{}{} \right. \nonumber \\  
                         &   &  \left( \frac{t'}{(E_{1}+E_{3})(E_{2}+E_{4})(E_{3}+E_{4})} - \frac{\sin(E_{1}+E_{3})t'}{(E_{1}+E_{3})^{2}(E_{1}-E_{4})(E_{1}-E_{2}+E_{3}-E_{4})} \right. \nonumber \\
                         &   & \left. -\frac{\sin(E_{3}+E_{4})t'}{(E_{3}+E_{4})^{2}(E_{1}-E_{4})(E_{2}-E_{3})} + \frac{\sin(E_{2}+E_{4})t'}{(E_{2}+E_{4})^{2}(E_{2}-E_{3})(E_{1}-E_{2}+E_{3}-E_{4})} \right) \nonumber \\
                         &   & \left. + n(E_{3})n(E_{4}) \left(\mbox{same terms with $E_{3,4}\rightarrow -E_{3,4}$}\right)\frac{}{} \right].
\end{eqnarray}
The other terms in Eq.~(\ref{eq:Asymmetry_H4}) can be done in a
similar way and they all have similar structures (with different
arrangements of energies in the sines and denominators).  There are
$12\times 4\times 2 = 96$ terms similar to
Eq.~(\ref{eq:Asymmetry_perturbative_5}) in the final result for the
perturbative lepton asymmetry.  Because of its size (there are no
obvious simplifications) and since it is not very instructive, we do
not write the full expression of the perturbative lepton asymmetry
here.

Even at the level of indivitual terms we see that
Eq.~(\ref{eq:Asymmetry_perturbative_5}) has the features expected from
the first two Sakharov conditions.  First it is clear that
Eq.~(\ref{eq:Asymmetry_perturbative_5}) is zero when $\eta = 0$,
$\epsilon = 0$ or $\delta=0$ (i.e. no $CP$ violation).  It also
vanishes when the sterile neutrino masses are degenerate (i.e. no
lepton number violation).  To see that, note that each terms in
Eq.~(\ref{eq:Asymmetry_perturbative_5}) is paired up with a similar
term with $1R\leftrightarrow 2R$ and $F_{\alpha 1}\leftrightarrow
F_{\alpha 2}$ (cf. Eq.~(\ref{eq:Asymmetry_perturbative_2})).  The
asymmetry is thus proportional to $E_{1}-E_{2}$ and vanishes when the
two energies are equal (i.e. when $\Delta M = 0$).

\subsection{Comparison Between Exact and Perturbative Solutions}
\label{Sec:Comparison}

The exact and perturbative computations of the asymmetry start with
the same initial condition and the dynamics is dictated by the same
Lagrangian.  They should therefore give the same result up to some
time where secular terms become important.  To estimate this time, it
is not possible to use the considerations of
Sect.~\ref{Sec:Validity_perturbation_theory} because there is no
interaction (thus no thermalization) and spacetime is not expanding. 
We use instead the following argument.

The toy model considered in Sect.~\ref{Sec:Application} is quadratic
in the field, thus the asymmetry production should be oscillatory. 
Looking at the exact~(\ref{eq:Asymmetry_exact_final}) and
perturbative~(\ref{eq:Asymmetry_perturbative_5}) solutions, we note
that both solutions are sums of oscillatory functions with different
frequencies.  The ``exact'' and ``perturbative'' frequencies are given
by $\tilde{\omega}_{ij,\pm} = (\tilde{E}_{i}\pm \tilde{E}_{j})$ and
$\omega_{ij,\pm} = (E_{i}\pm E_{j})$.  Since the masses are different
in the exact and perturbative cases, the frequencies are also
different.  This implies that the two solutions develop a phase
difference over time.  This phase difference is secular.  Thus even in
a non-interacting theory, secular terms are present because of the
building up of phase diffence between solutions.

The exact and perturbative solutions should agree when this phase
difference is small.  We estimate this phase difference in the
following.  The frequencies can be approximated as $\omega_{ij,\pm}
\approx |\vec{p}|(1\pm 1) + (m_{i}^{2}\pm m_{j}^{2})/2|\vec{p}|$ (we
assume that $m < |\vec{p}|$ here).  The criterion for the smallness of
secular terms is thus:
\begin{eqnarray}
\label{eq:Criterion_toy_model}
\left|\omega_{ij,\pm}-\tilde{\omega}_{ij,\pm}\right|t \ll 1 & \;\;\Rightarrow\;\; & t \ll \frac{2|\vec{p}|}{\left|(m_{i}^{2}\pm m_{j}^{2})-(\tilde{m}_{i}^{2}\pm \tilde{m}_{j}^{2}) \right|}.
\end{eqnarray}
Thus the set of masses that produces the largest frequency difference
gives the lowest time at which the two solutions should differ from
each other.  

\begin{figure}
\centering

\begin{tabular}{cc}
\includegraphics[width=0.5\textwidth]{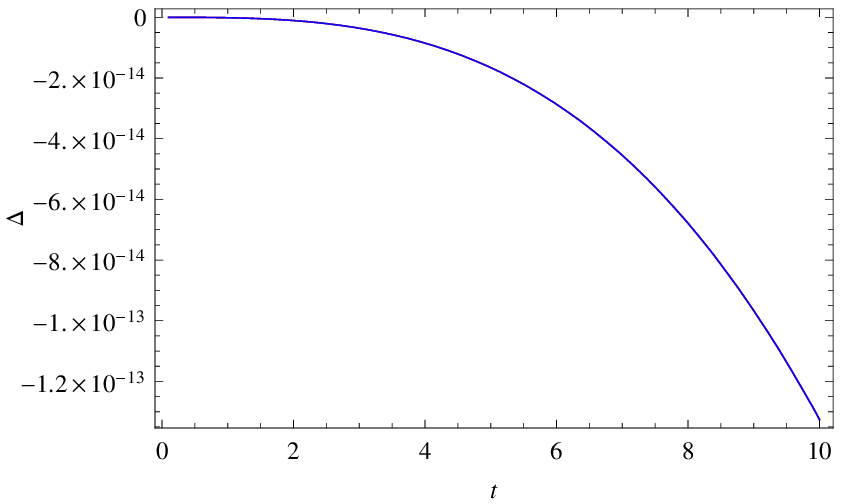} &
\includegraphics[width=0.5\textwidth]{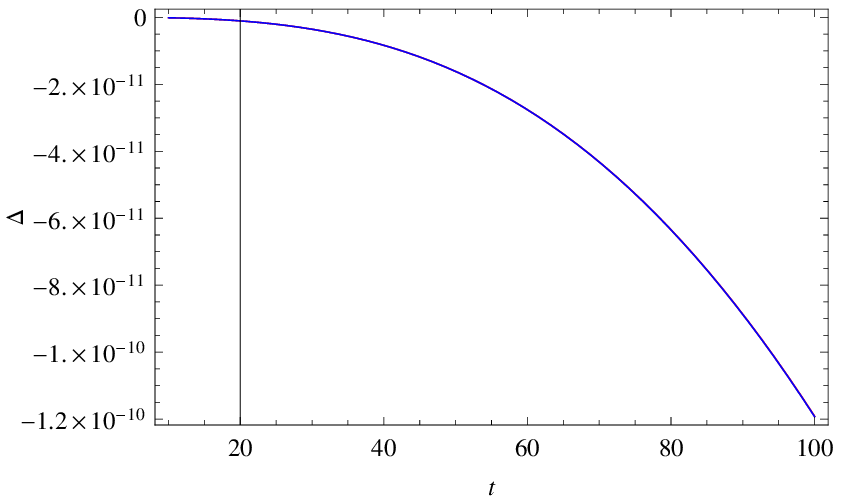}
\end{tabular}

\begin{tabular}{cc}
\includegraphics[width=0.5\textwidth]{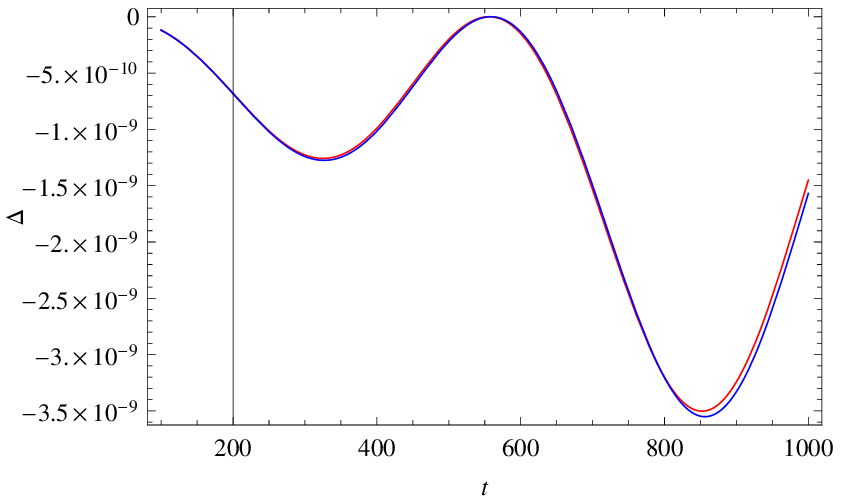} &
\includegraphics[width=0.5\textwidth]{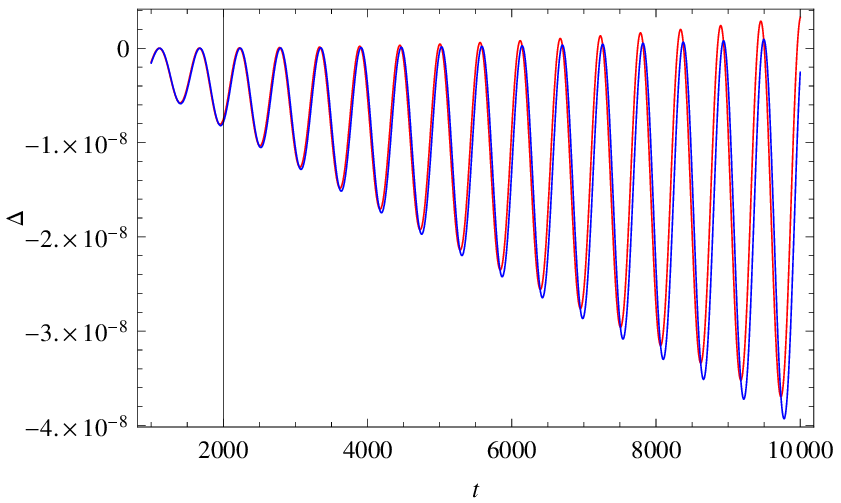}
\end{tabular}

\begin{tabular}{cc}
\includegraphics[width=0.5\textwidth]{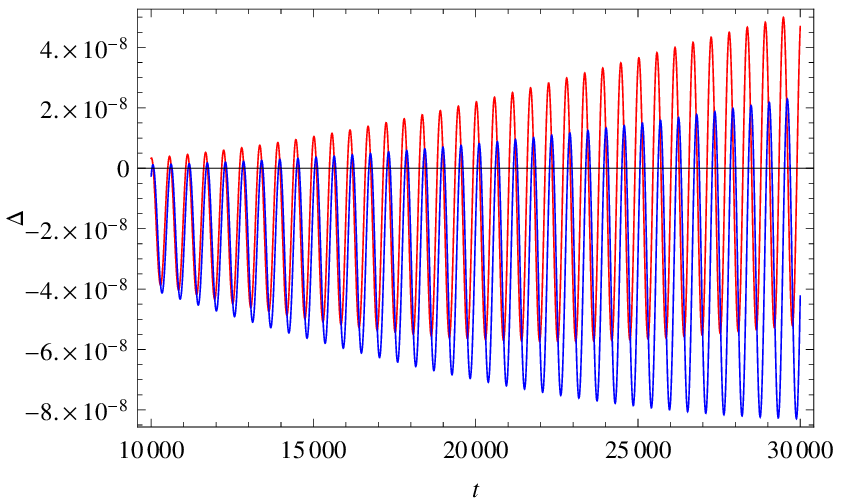} &
\includegraphics[width=0.5\textwidth]{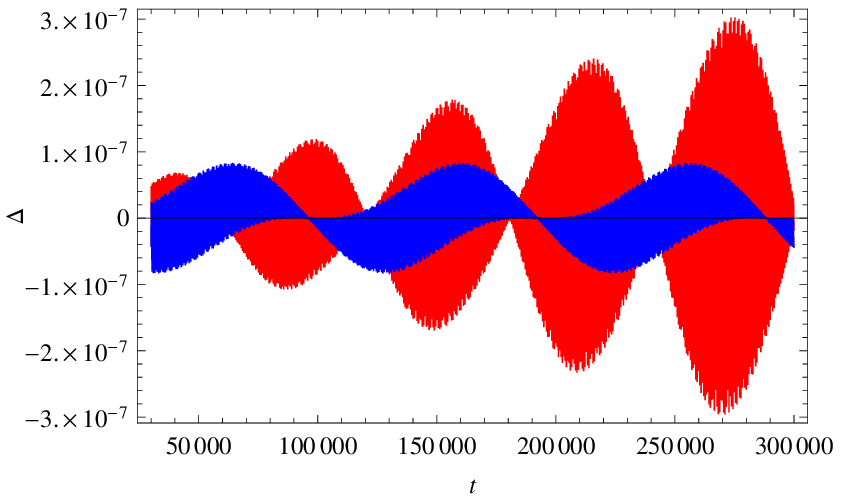}
\end{tabular}
\caption{Plots of active lepton asymmetry production (for one flavor) per unit of phase space as a function of time. The red and blue lines correspond to the exact and perturbative results, respectively.  Each plot covers a different time interval (time is expressed in GeV$^{-1}$).  We note that the exact and perturbative results agree very well up to $t\approx 6000 \approx t_{s}/3$, where we start to see small discrepancies that grow larger with time.}
\label{fig:Comparison}

\end{figure}

In the following we compare the exact
result~(\ref{eq:Asymmetry_exact_final}) and the complete perturbative
result (c.f. Eq.~(\ref{eq:Asymmetry_perturbative_5})) numerically.  We
plot both results for representative time intervals.  For simplicity
we plot the asymmetry per unit phase space volume and take a typical
momentum $p\sim T$.  The values of the parameters used are $f =
3\times 10^{-4}$, $v = 174$ GeV, $\epsilon = 0.8$, $\delta = 0.3$,
$\eta = 0.6$, $\kappa = 1.1$, $M = 1.5$ GeV, $T = 100$ GeV and $f =
10^{-7}$.  The results are shown in Fig.~(\ref{fig:Comparison}).


We can estimate the time $t_{s}$ at which secular terms become
important using the criterion~(\ref{eq:Criterion_toy_model}) and the
expression for the masses~(\ref{eq:Exact_masses}).  For the parameters
chosen here, the largest frequency difference is produced by
$\left|\omega_{12,\pm}-\tilde{\omega}_{12,\pm}\right|$:
\begin{eqnarray}
t_{s} & \ll & \frac{2|\vec{p}|}{\left|(m_{1}^{2}\pm m_{2}^{2})-
(\tilde{m}_{1}^{2}\pm \tilde{m}_{2}^{2}) \right|} \;\;\sim\;\; 
\frac{|\vec{p}|}{M\Delta M 
\left(\frac{1+\epsilon^{2}+\delta^{2}}{(\epsilon+\delta)\kappa}\right)},
\end{eqnarray}
which gives $t_{s}\sim 21000$ GeV$^{-1}$.  Note that all estimates are
roughly given by $t_{s} \sim |\vec{p}|/M\Delta M$.

\begin{figure}
\centering

\begin{tabular}{cc}
\includegraphics[width=0.5\textwidth]{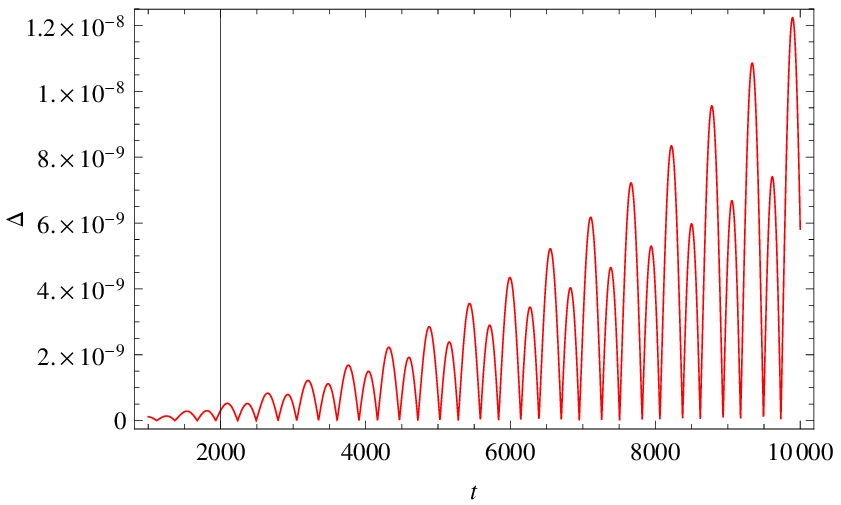} &
\includegraphics[width=0.5\textwidth]{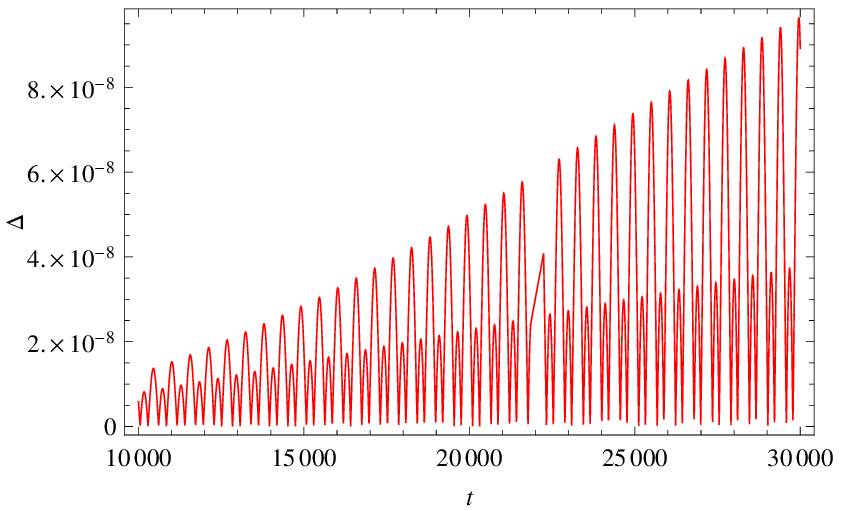}
\end{tabular}
\caption{Plot of the absolute error between the exact and perturbative results for two different time intervals.}
\label{fig:Absolute_error}

\end{figure}

As expected the two solutions follow each other nicely until $t\sim
6000\sim t_{s}/3$ where discrepancies appear.  Those discrepancies are
relatively small (roughly 10$\%$), but they steadily grow with time;
see Fig.~(\ref{fig:Absolute_error}) for plots of the absolute error
between the two results.  At $t\sim t_{s}\sim 21000$ the absolute
error becomes roughly as large as the absolute value of both
asymmetries.  We thus conclude that the exact and perturbative methods
agree with each other from small times to a time $t_{s}$ at which
secular terms are large and perturbation theory breaks down.


\section{Conclusion}

First principles computations of the lepton (baryon) asymmetry are
difficult because some degrees of freedom need to be
out-of-equilibrium in order to get a non-zero result, and the
treatment of those degrees of freedom using quantum field theory is
unwieldy.  In this paper we have derived from first principles of
quantum field theory and statistical mechanics a (simple) formula that
could be used as a starting point for a perturbative computation of
the baryon asymmetry of the universe.  Our formalism is quite general
and can be applied to other models.  The only assumption entering into
our derivation is that perturbation theory must be valid; physically
this translates into the condition that the total number of
out-of-equilibrium degrees of freedom (sterile neutrinos in our case)
must remain small.  This last condition depends on the parameters of
the model under study.  

We have also tested this formula for the asymmetry on an exactly
solvable toy model.  We have confidence that the method works and that
it can be applied to a more complicated model involving damping.  The
application of this formalism to the $\nu$MSM and the study of its
phenomenology is work in progress.

\begin{acknowledgments}
The authors would like to thank T. Asaka, T. Hambye and J. Louis 
for useful comments and discussions.  This work was supported in part by the Swiss National Science Foundation.
\end{acknowledgments}

\bibliography{baryon_asymmetry_bibliography}

\end{document}